%% file: main.tex
\documentclass[sn-mathphys,Numbered]{sn-jnl}
\usepackage[T1]{fontenc}

\usepackage{lipsum}
\usepackage{graphicx}%
\usepackage{multirow}%
\usepackage{amsmath,amssymb,amsfonts}%
\usepackage{amsthm}%
\usepackage{mathrsfs}%
\usepackage[title]{appendix}%
\usepackage{xcolor}%
\usepackage{textcomp}%
\usepackage{manyfoot}%
\usepackage{booktabs}%
\usepackage{algorithm}%
\usepackage{algorithmicx}%
\usepackage{algpseudocode}%
\usepackage{listings}%
\usepackage{tikz}
\usepackage{empheq}

\usepackage{ifpdf}

\ifpdf
  \DeclareGraphicsExtensions{.eps,.pdf,.png,.jpg}
\else
  \DeclareGraphicsExtensions{.eps}
\fi

\graphicspath{{Figure/}} 
\usepackage{macros}

\newcommand{\reviewerFour}[1]{{\color{black}{#1}}}
\newcommand{\ed}[1]{{\color{black}{#1}}}

\begin{document}

\title[Mathematical modelling and computational reduction of molten glass fluid flow in a furnace melting basin]{Mathematical modelling and computational reduction of molten glass fluid flow in a furnace melting basin}


\author[1]{\fnm{Francesco} \sur{Ballarin}}\email{francesco.ballarin@unicatt.it}

\author*[2]{\fnm{Enrique} \sur{Delgado \'Avila}}\email{edelgado1@us.es}

\author[3]{\fnm{Andrea} \sur{Mola}}\email{andrea.mola@imtlucca.it}

\author[4]{\fnm{Gianluigi} \sur{Rozza}}\email{grozza@sissa.it}

\affil[1]{\orgdiv{Department of Mathematics and Physics}, \orgname{Universit\`a Cattolica del Sacro Cuore}, \orgaddress{\city{Brescia}, \country{Italy}}}

\affil*[2]{\orgdiv{Departamento de Ecuaciones Diferenciales y An\'alisis Num\'erico}, \orgname{Universidad de Sevilla}, \orgaddress{\postcode{41080}, \city{Seville}, \country{Spain}}}

\affil[3]{\orgdiv{MUSAM Lab}, \orgname{Scuola IMT Alti Studi Lucca}, \orgaddress{\street{Piazza S. Ponziano, 6}, \city{Lucca}, \postcode{55100}, \country{Italy}}}

\affil[4]{\orgdiv{mathLab, Mathematics area}, \orgname{SISSA, International School for Advances Studies}, \orgaddress{\street{via Bonomea 265}, \city{Trieste}, \postcode{34136}, \country{Italy}}}


\abstract{In this work, we present the modelling and numerical simulation of a molten glass fluid flow in a furnace melting basin. We first derive a model for a molten glass fluid flow and present numerical simulations based on the Finite Element Method (FEM). We further discuss and validate the results obtained from the simulations by comparing them with experimental results. Finally, we also present a non-intrusive Proper Orthogonal Decomposition (POD) based on Artificial Neural Networks (ANN) to efficiently handle scenarios which require multiple simulations of the fluid flow upon changing parameters of relevant industrial interest. This approach lets us obtain solutions of a complex 3D model, with good accuracy with respect to the FEM solution, yet with negligible associated computational times.}


\maketitle

\section{Introduction}

Efficiently addressing Computational Fluid Dynamics (CFD) problems arising from industrial problems  with an appropriate numerical model in an affordable computational time is typically very challenging. While every application often comes with its own set of challenges, a common one is that the complexity of industrial mathematical models makes them impossible to solve in real time, especially in those cases where the model itself depends on several parameters. This work aims to exemplify the application of a non-intrusive reduced order modelling technique to a specific industrial application, namely the fluid flow of molten glass.

In order to achieve this goal, first of all we will present a mathematical model for molten glass fluid flow in a furnace melting basin. We will discuss in particular which physics are the most relevant for the application at hand, and therefore must be incorporated by means of appropriate partial differential equations in the model; the remaining physical contributions will be included instead by means of forcing terms or boundary conditions\reviewerFour{. After this, we will introduce a post-processing stage needed to compute the performance parameters desired by design engineers and furnace operators}. The molten glass is assumed to be a viscous fluid, which is heated in two ways: by methane combustion and by electric booster heating. The two heat sources will be modeled differently: on one hand, the heat flux due to the methane combustion flame is incorporated in the energy of the system by means of a boundary condition. On the other hand, a partial differential equation will be set up to model the booster heating, by considering an electrical potential equation. Thus, we are dealing with a (non-linear) multiphysics system, where the velocity, pressure and temperature of the fluid are coupled with the electric potential produced by the boosters. The resulting model is more complete than the one presented by previous authors, who discuss more basic models for molten glass, without the consideration of boosting heating \cite{Abbassi2008, Ma2018, Viskanta2006, Viskanta2006P2,  Shelby2007, Viskanta1994, Xiqi1986}. Even though modelling of booster heating has already been presented in a few previous works \cite{Chang2002, Choudhary1985, Choudhary2010, Curran1973, Giessler2009}, most of the simulations performed therein have been made by Finite Difference Methods (FDM). Here we consider instead 
 the Finite Element Method (FEM), since it allows us to consider more complex geometries than those previously discussed with FDM, and enables us to perform local refinement close to specific regions of interest in the domain. The resulting FEM model will then be validated against experimental measurements.

While the proposed FEM model \ed{allows us} to obtain reliable simulations against experimental data, the CPU time required to run each simulation results in a computational model which is too expensive to be queried in any practical scenario, which ideally would require a real time response (e.g., to monitor the furnace conditions and take appropriate action in case of undesired situations). To this end, we identify five parameters that \ed{mimic} a change in operating conditions in the furnace: one parameter corresponds to the amount of energy coming in the system by the methane combustion, and the other four correspond \ed{into} the voltage of the boosters. Thus, we are parameterizing the energy that is incorporated to the system.

In literature, many authors have proposed the use of Reduced Order Models (ROM) in order to reduce the computational time \cite{Quarteroni2014, QuarteroniROM, Quateroni2011}. Nowadays, two families of ROMs are mainly studied: the intrusive ROM and the non-intrusive ROM. The classic intrusive ROMs consist \ed{of} solving the reduced problem by performing a Galerkin projection onto the reduced spaces to compute the value of the modes weights, while the most recent non-intrusive ROMs approximate the modes weights, that only \ed{depend} on the parameter, by interpolation or regression techniques.

In order to achieve a real time visualization of the furnace state, we rely on a non-intrusive ROM based upon Artificial Neural Networks (ANNs) \cite{Chen2021, Hesthaven2018} for the molten glass fluid flow model. During a training phase, we construct the reduced spaces by a Proper Orthogonal Decomposition (POD). During the evaluation phase, we seek a reduced order solution of the problem as linear combination of the POD modes, where an ANN is queried to produce the weights to be employed in the combination. On one hand, by foregoing the use of an intrusive ROM we do not achieve certification of the error committed by the ROM \cite{Ballarin2020, ChaconRebollo2017,  POD1, Manzoni2014, Novo2021}; on the other hand, since we do not have to consider a Galerkin projection of the model equations onto the reduced spaces, this non-intrusive ROM lets us compute on-line solutions for this model that has highly non-linear terms, avoiding the use of techniques such as the EIM or DEIM \cite{DEIM2010, Barrault2004}. Non-intrusive ROMs \cite{Chen2021, Hesthaven2018, Pichi2021} are being increasingly considered in recent years when handling complex problems, and they have proven to be able to solve almost in real-time realistic industrial problems, since we avoid the treatment of the non-linear terms of the model equations. 

\ed{Other non-intrusive techniques can be addressed in this context. In \cite{Ivagnes2023, Tonicello2024} authors consider, in addition to ANN, the Radial Basis Functions (RBF) technique for the reconstruction of the POD coefficients, applied to CFD. In those works, authors compare the error between both techniques for the reconstruction of the POD coefficients, showing that the error of the POD-NN and POD-RBF are comparable.}

\reviewerFour{To summarize, this work presents an industrial application, which results in a nonlinear PDE problem coupling thermal, electrical and fluid dynamic fields in a complex domain. The spatial discretization has been successfully tackled with with Finite Elements Method, which resulted in a significantly lower numbers of the degrees of freedom of the numerical problem, with respect to the ones that would have been obtained with either Finite Differences or Finite Volumes methods typically used in these applications. To maximise the information provided to design engineers and furnace operators, a post-processing step has been devised to compute the motion of air bubbles which are released by the melting glass and might reduce the final product quality if present at the throat upon extraction from the furnace. Given its elevated computational cost, such a complicated problem represents an ideal --- albeit challenging --- test bench for model order reduction techniques. In particular, the goal of providing real time predictions to furnace operators has been achieved by means of  a non-intrusive ROM combining POD for the computation of modal basis and Artificial Neural Networks (ANNs) for the computation of the basis coefficients. Documenting the most critical aspects and the results obtained with the academic methods described when put to test on the present industrial problem represent critical information for the community. \cite{HijaziPCE2020} }

The structure of this paper is as follows. In Section 2, we present the problem modelling, with its mathematical formulation where the boundary conditions are also presented. In Section 3, we present the Finite Element discretization of the molten glass flow problem. In Section 4, is presented the validation of the Finite Element problem with experimental data, and a post-processing of the solution obtained. In Section 5, we present the \ed{Proper Orthogonal Decomposition Neural Network (POD-NN)} reduced order method and the numerical results of this model. 

\section{Fluid flow modelling of a glass furnace melting basin}

In the present section\ed{,} we discuss the modeling aspects of the container glass furnace melting basin simulations carried out in this work, namely a molten fluid glass flow. 

\subsection{Geometry of an industrial melting basin}

\begin{figure}[ht]
\centering
\centerline{
  \ifpdf
  \resizebox{1.0\textwidth}{!}{
    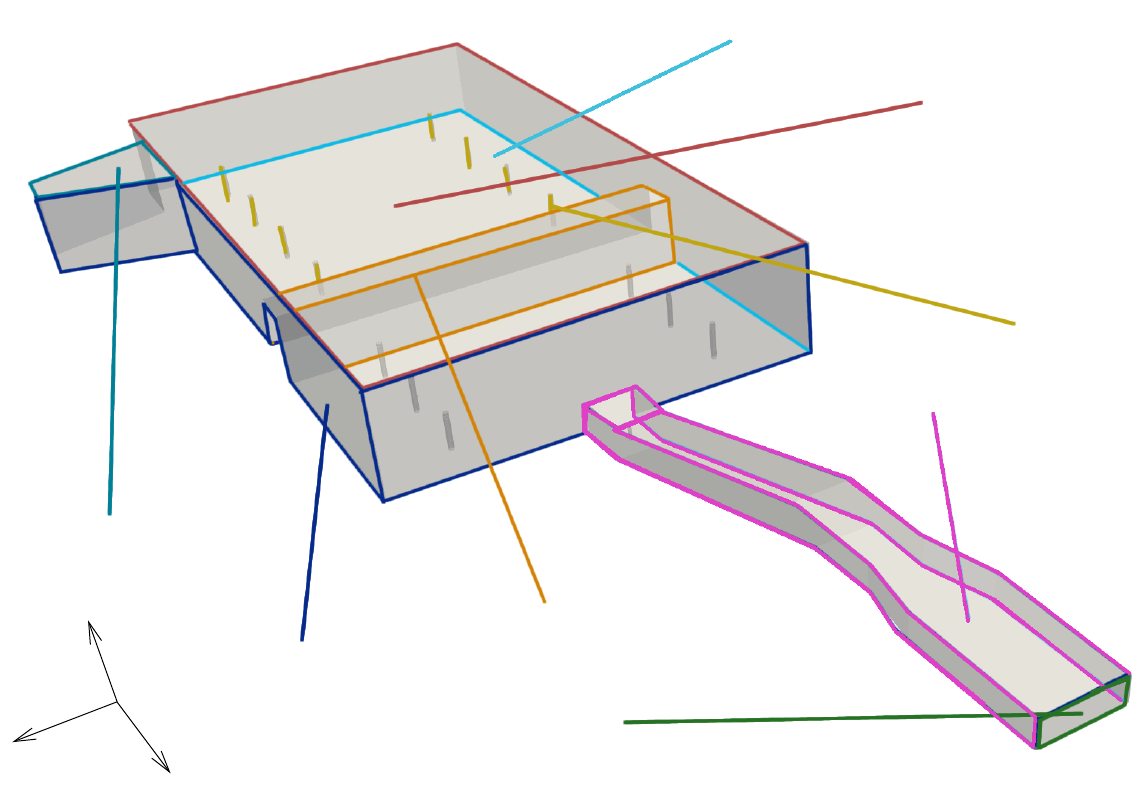
  }
  \else
  \resizebox{1.0\textwidth}{!}{
    \input{./Figure/boundaries.pstex_t}
  }
  \fi
}
  \caption{Industrial melting basin employed in this work. The boundaries indicated in the figure will be referred in this work as 
  \[
\begin{array}{l}
\Gamma_{in}:\text{Inflow of material.}\\
\Gamma_{top}: \text{Top of the furnace.}\\
\Gamma_{b_{i}}: \text{Surface of booster } i.\\
\Gamma_{w}: \text{The walls of the furnace.}\\
\Gamma_{bott}: \text{The bottom of the furnace.}\\
\Gamma_{wr}: \text{The wall for the recirculation.}\\
\Gamma_{duct}: \text{The walls of the throat.}\\
\Gamma_{out}: \text{The exit of the throat.}
\end{array}
\]
  \label{fig:boundaries}}
\end{figure}

The melting basin studied herein is depicted in Figure \ref{fig:boundaries}, and will be denoted in the following as the spatial domain $\Omega$. The melting basin reported in the Figure is a slightly simplified geometry of the actual basin employed by our industrial partners Bormioli Pharma Srl. We will provide a description of each part of the basin geometry next, yet refrain from reporting precise quantitative information on the geometry itself since such information is privileged and covered by \ed{the} intellectual property of the industrial partner. \ed{However, to allow for the qualitative reproduction of the results reported, we report that the overall length of the furnace considered is of the order of tenths of meters.}

The inflow of raw material occurs through the inflow section denoted as $\Gamma_{in}$ in Figure \ref{fig:boundaries}. The inflow section lies on top of a region, called the doghouse region in industrial practice, which in Figure \ref{fig:boundaries} has the shape of a trapezoidal prism and which goal is to collect all the material that enters in the basin. The material then moves to the actual furnace, \ed{whose} is modelled in Figure \ref{fig:boundaries} as the parallelepiped delimited by $\Gamma_{w}$, $\Gamma_{top}$ and $\Gamma_{bott}$. $\Gamma_{bott}$ and $\Gamma_{w}$ are the walls of the basin, placed respectively at its bottom and on its vertical sides. $\Gamma_{top}$ is not an actual physical wall, but the free surface that the molten glass assumes during the melting process; for reasons that will be discussed in the following, it is reasonable in our model to assume such \ed{a} surface to be flat. Furthermore, eight pairs of boosters are installed on the bottom surface; the surface of each booster is denoted by $\Gamma_{b_{i}}$, for $i = 1, \hdots, 8$. Finally, a barrier $\Gamma_{wr}$ is placed towards the end of the furnace to create a sudden contraction in the basin, with the goal of enhancing the recirculation of the molten glass before extraction of the finished product. The throat, represented by the elongated parallelepiped delimited by $\Gamma_{duct}$ and $\Gamma_{out}$ in Figure \ref{fig:boundaries}, constitutes the channel downstream of the furnace by which the finished product is extracted; in particular, the surface of the throat is denoted by  $\Gamma_{duct}$ and acts as a physical wall, while the outflow section of the finished product by $\Gamma_{out}$. The different boundaries of the domain $\Omega$ are indicated in Figure \ref{fig:boundaries}.

\subsection{Rationale and desiderata for the fluid flow model}
The main goal of the simulation tool developed herein is to model physical processes occurring in the melting basin in order to assess how furnace design parameters affect the quality of the molten glass extracted from the furnace at the end of the melting process. The main aspect of glass quality typically monitored in industrial practice is the presence of air and gas bubbles within the molten glass. In fact, such bubbles could be visible in the final product if they were still present in the fluid glass at the moment when the material is eventually molded into the desired containers, and thus they would become an undesired flaw.

\ed{The presence of bubbles within the fluid glass \ed{is} partly due to the progressive melting of the \emph{air-filled} granular raw material lying on the basin surface in the doghouse region, and partly due to chemical reactions occurring in the basing and releasing gas products. Thus, in principle a suitable model for the application at hand should be able to reproduce, along with fluid dynamic and thermal fields, the melting mechanism of granular material, as well as the chemical reactions leading to gas release in the fluid. However, given the fact that gas and air bubbles \ed{released} in the molten glass \ed{are} quite difficult to prevent, many of the design solutions and control parameters available to engineers and technicians focus instead on obtaining glass residence times in the furnace long enough for the air and gas to be released through the free surface of the basin. The presence of walls and electrodes (or boosters) which introduce heat through electric current are among the solutions put in place so as to create and control convective and recirculating flow patterns that increase the glass residence time.}

Therefore, in order to test the effects of new designs and control parameters, in this work we concentrate our efforts \ed{on} devising a model that results in the accurate reproduction of the molten glass motion in the basin through the mechanisms affecting the electric potential, thermal and fluid dynamic fields, but not the production of air bubbles. However, as it will be detailed in the following sections, once the fluid velocity field has been obtained, we will evaluate the motion of small air bubbles by means of a \ed{post-processing} step, under the assumption that their presence does not significantly affect the fluid dynamic field.

\subsection{Mathematical model}

For the aforementioned reasons, and considering the low Reynolds number of the flow at hand, the governing equation adopted in this work \ed{is} the transient three-dimensional incompressible Navier-Stokes equations, which describe the motion of a viscous Newtonian fluid. To include thermal convection effects even in \ed{the} presence of an incompressible fluid model, we resort to Boussinesq approximation for the buoyancy terms. Finally, the heat source generated by  boosters is considered in the model as a \ed{one-way} coupling term between the electric potential field and the thermal field.

Let $\Omega \subset \mathbb{R}^3$ be the bounded polyhedral computational domain introduced in section 2.1, and $[0, T]\subset\mathbb{R}$ the time interval. The proposed mathematical model is governed by the following partial differential equations
\reviewerFour{
\begin{subequations}\label{Boosting-model}
\begin{empheq}[left=\empheqlbrace]{align}
&\rho \partial_t\textbf{u} + \rho (\textbf{u}\cdot\nabla\textbf{u})-\nabla\cdot(\mu\nabla\textbf{u})+\nabla p - \rho \textbf{g}\beta (\theta-\theta_0)=\textbf{0} &\text{in }&\Omega\times[0,T], \\ 
&\nabla\cdot\textbf{u}=0 &\text{in }&\Omega\times[0,T],\\
&\rho c_p \partial_t\theta+\rho c_p (\textbf{u}\cdot\nabla\theta)- \nabla\cdot(k_{eff}\nabla\theta)=\sigma|\nabla\phi|^2&\text{in }&\Omega\times[0,T],\\
\label{Boosting-model:sigma}
&-\nabla\cdot(\sigma\nabla\phi)=0 &\text{in }&\Omega\times[0,T].
\end{empheq}
\end{subequations}
}
The unknowns of the model are the fluid velocity $\textbf{u}$ [$m/s$], the fluid pressure $p$ [$\ed{Pa}$], the fluid temperature $\theta$ [$K$]\ed{,} and the electric potential $\phi$ [$V$]. \ed{Here, $\textbf{g}$ denotes the gravity.}
The expressions for the physical properties appearing in \eqref{Boosting-model} are summarized in Table \ref{Tab:Forno2}, where $\rho$, $c_p$, $\mu$, $\beta$,  $k_{eff}$, and $\sigma$ are, respectively, the density, the specific heat, the viscosity, the thermal expansion coefficient, the thermal conductivity, and the electrical conductivity of the fluid.

We remark that viscosity, thermal and electrical conductivity show dependence on the local temperature. \ed{Their expressions as reported in Table \ref{Tab:Forno2} are based upon the pyshical properties considered in \cite{Choudhary1985}.}
\reviewerFour{We point out that the expression for the viscosity,

\begin{equation}
\mu(\theta) = \exp\left(\dfrac{10425}{\theta-500}-6.0917\right) \left[\frac{\text{kg}}{\text{m}\,\text{s}}\right]
\end{equation}}
presents a vertical asymptote at $\theta_a=500K$, which would prevent to reach low temperatures for the material, as the viscosity would become infinite as the temperature approaches $\theta_a$. Considering that the viscosity model reported in Table \ref{Tab:Forno2} is no longer valid for low temperatures as it would be nonphysical for the viscosity to reach an arbitrarily large value, we set a threshold at $\theta_c=973K$, and modify the viscosity expression as follows: if $\theta \geq \theta_c$, than the viscosity law is modelled as in Table \ref{Tab:Forno2}, otherwise we assume a constant viscosity equal to the one obtained at $\theta_c$. 
Moreover, the power at the electrodes is modeled by a source term in the energy equation, as $\sigma(\theta)|\nabla \phi|^2$. This corresponds with the heat generated by Joule effect by the boosters. \reviewerFour{Taking a look at Equations \ref{Boosting-model}, it would be in principle possible to consider a one-way coupling between electric potential and temperature field, $\sigma$ showed only marginal dependence on $\theta$. In this work, variations of electrical conductivity as a function of temperature (\textit{cf. }\cite{Choudhary1985}) are expressed by
\begin{equation}
    \sigma(\theta) = \exp(7.605-7200/\theta)\  [1/(\Omega\, m)].
\end{equation}
In the temperature range $T\in[973^\circ K; 1900^\circ K]$ observed in the furnace, the value of $\sigma$ then varies between $1.2\  \Omega^{-1}m^{-1}$ and $44.7\  \Omega^{-1}m^{-1}$. Given this extreme temperature sensitivity of the electrical conductivity, Equation \ref{Boosting-model:sigma} is strongly affected by temperature changes, which can only be tackled considering  a two way coupling between the fields.}






\begin{table}[ht]
\centering
\begin{tabular}{l|l|l|l}

\hline
\hline

$\beta$ & Thermal expansion & $7.5\cdot10^{-5}$ & $[1/K]$\\
$\theta_0$ & Reference temperature & $1617$ & $[K]$\\
$\rho$ & Density & $2250$& $[kg/m^3]$\\
$c_p$ & Specific heat & $1381$& $[J/(kg\,K)]$\\
$\mu$ & Viscosity & exp$\left(\dfrac{10425}{\theta-500}-6.0917\right)$ & [$kg/(m\,s)]$\\
$k_{eff}$ & Thermal conductivity & $1.73 + 2.5\cdot10^{-8}\theta^3 $ & $[W/(m\,K)]$\\
$\sigma$ & Electrical conductivity & exp$(7.605-7200/\theta)$ & $[1/(\Omega\, m)]$\\
\hline
\hline

\end{tabular}
\caption{Physical properties of the model.}\label{Tab:Forno2}
\end{table}


\subsection{Boundary Conditions}
We present next the boundary conditions accompanying the partial differential equations \eqref{Boosting-model}. We discuss separately boundary conditions for velocity/pressure, temperature and electric equations.

\subsubsection{Velocity and pressure boundary conditions}
For the boundary conditions associated \ed{with} the Navier-Stokes momentum equation, we consider a prescribed inlet velocity in the normal direction to $\Gamma_{in}$. Assuming the normal to $\Gamma_{in}$ in Figure \ref{fig:boundaries} to be aligned with the vertical direction $\mathbf{e}_y$, the velocity at the inlet is given by:
\begin{equation}\label{eq:inlet_bc}
\uk_{in} = u_y \mathbf{e}_y\quad \text{on } \Gamma_{in}\ed{.} 
\end{equation}
Since our simulations start from an empty furnace, we consider a \ed{time-modulated} function $u_y = u_y(t)$, in which during the first 72 hours we gradually increase the quantity of material introduced in the furnace, after which the desired inflow velocity is reached and kept constant. Thus, the function $u_y(t)$ in (\ref{eq:inlet_bc}) is defined as
\begin{equation}
u_y(t)=\left\{\begin{array}{ll}
\left[\dfrac{1}{2}\sin{\left(\dfrac{\pi t}{72\cdot 3600} -\dfrac{\pi}{2}\right)} +\dfrac{1}{2}\right]\bar{u}_y\quad& \text{if } t\le 72\cdot 3600 s\smallskip\\
\bar{u}_y& \text{if } t> 72\cdot 3600 s,
\end{array}\right.
\end{equation}
where\reviewerFour{, in order to respect no slip boundary conditions at the surrounding walls, }$\bar{u}_y$ is a parabolic profile scaled in such a way to obtain a mass flow of 13.7 tons per day.

With respect to the side walls, bottom, the wall for the recirculation, \ed{throat}, and boosters walls, the velocity is set to zero (no-slip condition), thus,
\[
\uk=\textbf{0}\quad \text{on } \Gamma_w\cup\Gamma_{bott}\cup\Gamma_{wr}\cup\Gamma_{duct}\cup\Gamma_{b_{i}}.
\]

For the top of the furnace, since there are no walls on the top boundary, we have considered slip boundary conditions for the velocity, that is

\[
\uk\cdot\nk=0 \quad \text{on }\Gamma_{top}.
\]

Finally, at the outlet, we consider the outflow conditions, given by
\[
\dnormal{\uk}-p\nk=\textbf{0} \quad \text{on }\Gamma_{out},
\]
\ed{where we are denoting by $\textbf{n}$ the outward normal vector, and by $\partial_n$ the normal derivative.}

\subsubsection{Thermal boundary conditions}

Concerning the temperature field boundary conditions, we have considered a constant temperature on  the inlet boundary. Thus, we consider
\[
\theta=\theta_{in} \quad \text{on }\Gamma_{in}.
\]
In particular, we consider that the temperature on the inlet boundary is fixed at $\theta_{in}=1073K$. 

On the top boundary, we consider a heat flux boundary condition, given by
\[
k_{eff}\dnormal{\theta} = q_{top}\quad \text{on }\Gamma_{top}.
\]
where $q_{top}$ is a spatial function for the thermal boundary condition on the top boundary. This function is 
\begin{equation}\label{eq:BC_combustion}
    \reviewerFour{q_{top}=\max\{Q(z-z_{min})(z-z_{max})(x-x_{max})(x-x_{min}),0\},}
\end{equation}

\reviewerFour{where $x_{min},x_{max},z_{min}, \text{and}\ z_{max}$ are the coordinates of the boundaries of the basin top surface hit by the flame}, and $Q$ is computed such that the total heat flux is $269662kW/(m^2s)$. This flux intends to model the heat flux of methane combustion flame as a parabolic function along both $x$ and $y$ coordinates.\reviewerFour{ We point out that making use of different values of $x_{min},x_{max},z_{min}, \text{and}\ z_{max}$ is is also possible to deform the shape of the flame.}

On the side walls and on the bottom of the furnace\ed{,} we consider convection boundary conditions corresponding \ed{to} Robin boundary conditions. By supposing the external temperature to be $\theta_{ext} = 300K$, these conditions are given by
\[
k_{eff}\partial_n\theta=h_w (\theta - \theta_{ext})\quad \text{on }\Gamma_{w},
\]
\[
k_{eff}\partial_n\theta=h_{bott} (\theta - \theta_{ext})\quad \text{on }\Gamma_{bott},
\]
and
\[
k_{eff}\partial_n\theta=h_{duct} (\theta - \theta_{ext})\quad \text{on }\Gamma_{duct},
\]
with $h_w=6.123 W/m^2K$, $h_{bott}=0.531W/m^2K$\ed{,} and $h_{duct}=0.05 W/m^2K$ respectively. These heat transfer coefficients depend on the insulation material of each wall considered.

Finally, both on the outflow and the boosters we have considered adiabatic boundary conditions, that is
\[\partial_n\theta=0 \quad \text{on }\Gamma_{out}\cup\Gamma_{b_{i}}.\]
 
\subsubsection{Electric boundary conditions}
Concerning the electric boundary conditions, we consider a fixed potential en each electrode, that is

\begin{equation}\label{eq: BC_electricity}
  \phi=\tilde{V}_i\quad \text{on } \Gamma_{b_i} \quad i=1,\dots,8.  
\end{equation}

 These $\tilde{V}_i$ are imposed so as we have the same $\Delta V$ at each pair of electrodes. In particular, we consider that each pair of booster has \ed{$\tilde{V}_{2j-1}=V_j/2$ and $\tilde{V}_{2j}=-V_{j}/2$, $j=1,2,3,4$}, with $V_1 = 150.6V$,$V_2 = 132.4V$, $V_3=112V$, $V_4=124.6V$, corresponding from the first pair of boosters to the fourth, respectively.

Finally, we consider homogeneous Neumann boundary condition, on the rest of the boundary, i.e.,  
\[\partial_n\phi=0 \quad \text{on }
\Gamma_{in}\cup\Gamma_{out}\cup\Gamma_{w}\cup\Gamma_{top}\cup\Gamma_{wr}\cup\Gamma_{bott}\cup\Gamma_{duct}.\]

\subsection{Air bubbles motion as a \ed{post-processing}}


Once the glass velocity and thermal field are computed, it is possible to make use of the flow field to predict the motion of air bubbles in the molten glass. Assuming that the air bubbles are spheres with a radius $r$ that is smaller than the scales of the velocity field computed, the relative motion of the sphere in the flow field results equivalent to the simpler problem of a sphere moving in a fluid at rest. In addition, given the small relative velocity between the bubble and the fluid, we will assume that air bubbles\ed{'} motion can be modelled under Stokes flow and null inertial contribution. \reviewerFour{In such a case, the air bubble is subjected to the hydrostatic buoyancy force and to the fluid dynamic resistance, which in the case of Stokes flow has closed form. The force balance along the vertical direction for an air bubble of radius $r$ rising with relative speed $V$ with respect to the surrounding glass then reads

\begin{equation}
    \frac{4}{3}\pi r^3 g (\rho_g(\theta)-\rho_a(\theta))= 6\pi r \mu(\theta) V
\end{equation}
where $\rho_a(\theta)$ is the air density. $\rho_ga(\theta)$ is the glass density and $g$ the gravity acceleration. The equation results in an expression for the vertical velocity of the rising bubble

\begin{equation}
V=\frac{2}{9}\pi r^2 g \frac{(\rho_g-\rho_a)}{\mu},
\end{equation}
where the temperature dependency of the densities and of the glass viscosity is henceforth omitted to lighten the notation.  In this work, however, we used a slight modification of this formula, namely
\begin{equation}\label{eq:bubble_rise_vel}
V_b=\frac{g\Delta\rho r^2}{3\mu},
\end{equation} 
based on experimental evidence reported in \cite{Shelby2007}, where $\Delta\rho = \rho_g-\rho_a$. Equation~\eqref{eq:bubble_rise_vel} applies to bubbles with radius smaller than 1 mm. Making use of the glass density $\rho_g$ and viscosity $\mu$ values reported in Table~\ref{Tab:Forno2}, the Reynolds number $Re_{V_b}=\frac{2}{9}\frac{\Delta\rho^2}{\mu^2}gr^3$ obtained for 1 mm bubbles is significantly smaller than 1, which justifies Stokes flow assumptions. Thus, the experimental correction in equation~\eqref{eq:bubble_rise_vel} is more likely justified by a non perfectly spherical shape of the rising bubbles measured in the experiments. As said, the bubble rising displacement is the only relative motion between the air sphere and the surrounding fluid. For such reason}, the local air bubble velocity field is obtained using the local horizontal components of the fluid velocity, and obtaining the vertical component as the sum of $V_b$ and the local fluid velocity vertical component.

\reviewerFour{
\section{Numerical model for full order and reduced order solver}

In the present section, we will first describe the numerical solver based on Finite Element Method (FEM) used for the resolution of the mathematical model equations described in the previous section. After this, we will also provide details of the fully data driven POD-NN reduced order model implemented to accelerate the molten glass flow simulations.

\subsection{Finite element modelling}
To discretize the mathematical problem  \eqref{Boosting-model} by means of the Finite Element Method, we start defining the variational formulation of problem \eqref{Boosting-model}.
}
We consider the space of square integrable functions $L^2(\Omega)$ and the So\-bo\-lev space $H^1(\Omega)$, equipped with the standard norm. For simplicity in the notation, let us denote 
\[
\ed{\Gamma_{Dv} =\Gamma_w\cup\Gamma_{bott}\cup\Gamma_{wr}\cup\Gamma_{duct}\cup\Gamma_{b_{i}}.}
\]
Thus, we consider the following spaces for velocity, pressure, temperature and electric potential, respectively
\ed{
\[
W = \{\textbf{v}\in [H^1(\Omega)]^3 : \textbf{v}\cdot \textbf{n} = 0 \text{ on } \Gamma_{top}, \textbf{v}=0 \text{ on } \Gamma_{Dv}, \textbf{v} = \textbf{u}_{in} \text{ on } \Gamma_{in}\},
\]
\[
M = \{q\in L^2(\Omega)\},\quad \Theta = \{z\in H^1(\Omega) : z = \theta_{in} \text{ on } \Gamma_{in}\},
\]
and,
\[
\Phi = \{\omega\in H^1(\Omega) : \omega = \tilde{V}_i \text{ on } \Gamma_{b_{i}}\}.
\]

Moreover, we denote by $W_0, \Theta_0,$ and $ \Phi_0$, the spaces for velocity, temperature and electic potential with homogeneus boundary conditions, that is 

\[
W_0 = \{\textbf{v}\in [H^1(\Omega)]^3 : \textbf{v}\cdot \textbf{n} = 0 \text{ on } \Gamma_{top}, \textbf{v}=0 \text{ on } \Gamma_{Dv}\cup\Gamma_{in}\},
\]
\[
\Theta_0 = \{z\in H^1(\Omega) : z = 0 \text{ on } \Gamma_{in}\}, \quad
\Phi_0 = \{\omega\in H^1(\Omega) : \omega = 0 \text{ on } \Gamma_{b_{i}}\}.
\]
}
We consider the variational form of problem 
\eqref{Boosting-model} given by
\\[5mm]
$\text{Find } (\textbf{u}, p, \theta, \phi) \in W\times M\times\Theta\times\Phi, \text{such that} $\\[-10mm]

\begin{equation}\label{VariationalForm}
\left\{\begin{array}{ll}
(\rho \partial_t\textbf{u}, \textbf{v}) + (\rho (\textbf{u}\cdot\nabla\textbf{u}), \textbf{v})+ (\mu\nabla\textbf{u}, \nabla\textbf{v}) \smallskip\\
- (p, \nabla \cdot\textbf{v}) - (\rho \textbf{g}\beta (\theta-\theta_0), \textbf{v})= \textbf{0}   & \forall \textbf{v}\in W_0 \medskip\\
(\nabla\cdot\textbf{u}, q) =0 & \forall q\in M \medskip\\
(\rho c_p \partial_t\theta, z)+ (\rho c_p (\textbf{u}\cdot\nabla\theta, z) + ((k_{eff}\nabla\theta),\nabla z)= (\sigma|\phi|^2 , z) & \forall z\in \Theta_0 \medskip \\
(\sigma\nabla\phi, \nabla \omega )= 0 & \forall \omega\in \Phi_0.
\end{array}\right.
\end{equation}
From this variational formulation, we can define the Finite Element problem. Let us consider a regular mesh $\Th$ composed \ed{of} tetrahedrons. Given an integer $n\ge0$ and an element $K\in \Th$, we denote by $\mathbb{P}_n(K)$ the Finite Element space given by Lagrange polynomials of degree less than, or equal to $n$ defined on each element $K$ of $\Th$. Thus, we define
\[
Y^n = \{v_h \in C^0(\bar{\Omega}) \text{ such that } v_{h|_K}\in \mathbb{P}_n(K) \},
\]
\reviewerFour{where $v_{h|_K}$ denotes function $v_h$ evaluated on element $K$ of $\Th$.} Moreover, we define the bubble element, given by
\[
B^n = \{v_h \text{ such that } v_{h|_K}\in \mathbb{P}_n(K) \cap H_0^1(K)\}\ed{.}
\]

Let $W_h = W \cap ([Y^1]^3 \oplus [B^4]^3)$, $M_h = M \cap Y^1$, $\Theta_h = \Theta \cap Y^2$ and $\Phi_h = \Phi \cap Y^1$ be four finite dimensional subspaces of $W, M, \Theta, $ and $\Phi$, respectively. Note that we are considering for the velocity-pressure pair the so called mini-element of $\mathbb{P}_1$ + bubble for velocity and $\mathbb{P}_1$ for pressure since this pair of Finite \ed{Elements} is stable (see e.g. \cite{Arnold1984}). For temperature, we are considering $\mathbb{P}_2$ Finite Element since is the field of interest. For simplicity in the notation, let us denote $X_h=W_h\times M_h\times\Theta_h\times\Phi_h$.

For the temporal derivative terms in \eqref{VariationalForm}, we consider a fully implicit standard Backward Differentiation Formula (BDF) approximation (see e.g. \cite{Quarteroni2014}) of order one and of time step size $dt$. 
Thus, the scheme for solving numerically \ed{the} problem \eqref{VariationalForm} reads

\begin{equation}\label{eq:FE_problem}
\left\{
\begin{array}{ll}
\text{Given }(\textbf{u}^n, p^n, \theta^n, \phi^n)\in X_h, \\\text{find } (\textbf{u}^{n+1}, p^{n+1}, \theta^{n+1}, \phi^{n+1}) \in X_h,
\text{such that} \bigskip \\ 
(\rho \dfrac{\textbf{u}^{n+1} - \textbf{u}^n}{dt}, \textbf{v}) + (\rho (\textbf{u}^{n+1}\cdot\nabla\textbf{u}^{n+1}), \textbf{v}) - (p^{n+1}, \nabla \cdot\textbf{v})  \smallskip\\
+  (\mu(\theta^{n+1})\nabla\textbf{u}^{n+1}, \nabla\textbf{v}) - (\rho \textbf{g}\beta (\theta^{n+1}-\theta_0), \textbf{v})= \textbf{0}   & \forall \textbf{v}\in W_0 \medskip\\
(\nabla\cdot\textbf{u}^{n+1}, q) =0 & \forall q\in M \medskip\\
(\rho c_p \dfrac{\theta^{n+1}-\theta^{n}}{dt}, z) + (\rho c_p (\textbf{u}^{n+1}\cdot\nabla\theta^{n+1}, z) \\ 
+ \,((k_{eff}(\theta^{n+1})\nabla\theta^{n+1}),\nabla z)= \ed{(\sigma|\phi^{n+1}|^2 , z)} & \forall z\in \Theta_0 \medskip \\
(\sigma(\theta^{n+1})\nabla\phi^{n+1}, \nabla \omega )= 0 & \forall \omega\in \Phi_0
\end{array}
\right.
\end{equation}

In each iteration, we solve the nonlinear system \ed{through} the Newton method.

\reviewerFour{
\subsection{POD-NN reduced order model}\label{sec:POD-NN} 
}
In this section we present a non-intrusive POD for a parametrized PDE, based on a Neural Network. In particular, we consider as parameters the total amount of energy released on the furnace. As explained in previous sections, heat is introduced in two ways in the furnace system: the heat source relative to the methane combustion, and the heat produced by the boosting due to the Joule effect. Thus, the parameters considered are the total heat flux due to combustion in the top chamber (see equation \eqref{eq:BC_combustion}), and
the voltage (see equation \eqref{eq: BC_electricity}) of each of the four pairs of boosters. With this configuration, we are assuming that our parameterized PDE \ed{depends} on five parameters.

\ed{Considering a clasical intrusive Reduced Order Model based on POD and a Galerkin projection  \cite{POD1, POD_chapter} of problem \eqref{eq:FE_problem} is hard due to the fact that non-linearities present in the model. Those non-linearities would force us to employ, for example, a (Discrete) Empirical Interpolation Method (see e.g. \cite{Barrault2004, DEIM2010}), in order to forego employing a Galerkin projection. In order to avoid this, we consider a Neural Network in order to approximate the coefficients of the POD by regression.}

First of all, we recall the construction of the POD modes. Let us consider $\cD_{train} = \{\muk^1,\dots, \muk^P\} \subset \mathcal{D}$ a set of $P$ parameter values, and the corresponding set of snapshots for each field obtained by collecting the \ed{steady} state solutions corresponding to each parameter value. For simplicity of the notation, in the following we consider only the velocity field and its associated snapshot matrix $S_{\uk}=\{\uk(\muk^1), \dots,\uk(\muk^P)\}$ but, in the numerical results, we will display ROM results for all solution fields by proceeding analogously with \ed{what} is discussed below for the velocity field.

We construct the correlation matrix for each field, given by
\[
\mathbf{C}_{ij} = \dfrac{1}{P}(\uk(\muk^i), \uk(\muk^j))_W, \quad 1\le i,j\le P,
\]
with $(\cdot, \cdot)_W$ a scalar product on $W$. For the definition of the POD modes, we solve the following eigenvalue problem associated to the correlation matrix:
\begin{equation}\label{pb:eigenvalue}
\left\{
\begin{array}{l}
\text{Find }(\lambda_n, \vk_n)\in \R\times\R^P \text{ such that }\\
    \mathbf{C}\vk_n = \lambda_n\vk_n, \quad 1\le n\le P.
\end{array}   \right. 
\end{equation}

Solving this problem, we obtain $P$ eigenvalues sorted in descending order, $\lambda_1\ge \dots\ge \lambda_P$. With the $N$ largest eigenvalues, we define the POD modes as
\begin{equation}\label{eq:POD_modes}
\Psi^\uk_n = \D\sum_{k=1}^P(\vk_n)_k\uk(\muk^k), \quad 1\le n\le N,    
\end{equation}
where we are denoting by $(\vk_n)_k$ the $k$-th coefficient of the eigenvector $\vk_n\in\R^P$. Analogously, we can define the POD modes for pressure ($\Psi_n^p$), temperature ($\Psi_n^\theta$), and electric potential ($\Psi^\phi_n$). 

We define the Reduced Order spaces, for velocity, pressure, temperature and electric potential as following
\begin{equation}\label{POD_spaces}
\begin{array}{ll}
 W_N = \Span\{\Psi_n^\uk, \quad 1\le n\le N\},   &   M_N = \Span\{\Psi_n^p, \quad 1\le n\le N\},\\
  \Theta_N = \Span\{\Psi_n^\theta, \quad 1\le n\le N\},    &  \Phi_N = \Span\{\Psi_n^\phi, \quad 1\le n\le N\}.
\end{array}
\end{equation}

We consider a feed-forward Neural Network in order to build a regression model able to compute efficiently the value of the POD coefficients that better approximate the solution of problem \eqref{eq:FE_problem}. Some previous works on POD-NN can be found in \cite{Chen2021, Hesthaven2018, Pichi2021}. In that sense, we consider for each parameter value $\muk\in\cD$,  that the solutions of the \reviewerFour{POD-NN} problem are given by
\begin{equation}\label{eq:NN-POD_modes}
\begin{array}{cc}
\uk_{NN}(\muk) = \D\sum_{k=1}^N \alpha^\uk_k(\muk) \Psi_k^\uk, & p_{NN}(\muk) = \D\sum_{k=1}^N \alpha^p_k (\muk)\Psi_k^p,      \\
\theta_{NN}(\muk) = \D\sum_{k=1}^N \alpha^\theta_k(\muk) \Psi_k^\theta, & \phi_{NN} (\muk)= \D\sum_{k=1}^N \alpha^\phi_k (\muk)\Psi_k^\phi, 
\end{array}
\end{equation}
where the coefficients $\alpha^\uk_k$, $\alpha^p_k$, $\alpha^\theta_k$ and $\alpha^\phi_k$ are computed as follows (again, for simplicity only for the velocity field).
%
A Neural Network ROM applied to this problem consists in approximating a function $\pi : \cD\rightarrow\R^N$, that maps each parameter value $\muk\in\cD$ to the vector of POD coefficients:
\begin{equation}\label{map_pi}
\pi: \muk\in\cD \mapsto \alphak^{\uk}(\muk) = (\alpha_1^\uk(\muk), \dots, \alpha_N^\uk(\muk)).     
\end{equation}

This function $\pi$ is determined through a supervised learning approach, based on a training set given by the pairs $\D\{(\muk^k, \pi(\muk^k))\}_{k=1}^{P}$. \reviewerFour{ We remark that in the present work, we elected to train a separate NN for the modal coefficients of each field appearing in the original PDE. This choice is motivated by the fact that --- as will be further discussed in Section \ref{sec:ROMRes} --- we considered a limited amount of FOM solutions in the ROM training. Thus, the resulting NNs, characterized by smaller output space dimensions with respect to a that of a single NN for all fields, leads to more accurate predictions. On the other hand, a single NN for all fields would surely better reproduce the interdependence among the different fields and parameters involved in the simulations.
}
So, in this work, we have chosen a standard feed-forward Neural Network. We consider an input layer of dimension $M$ (the number of parameters considered in our problem), an output of dimension $N$ (the number of POD coefficients to compute \reviewerFour{for each field}), and $L$ inner layers, each compound of $L_N$ computing neurons. We set a learning rate $\eta$, the weighted sum as a propagation, the hyperbolic tangent as activation functions, and, in the last layer, the identity as output function. For the learning procedure, we consider a training set of $(2/3)P$ parameter values, and a validation set of $(1/3)P$ parameter values. The network is trained by minimizing a loss function, defined as the mean squared error between the POD-NN and FOM solutions.\\

\section{Validation and \ed{post-processing} of the finite element model}
In this section, we present the numerical results for the Finite Element problem presented in the previous section, its validation against experimental measurements, and its \ed{post-processing} to obtain the motion of air bubbles.

Since we are considering that the glass in the inlet is already melted, this heat flux is considered such that the energy consumed in the melting process is comes from the real heat flux by combustion. The total heat flux by combustion is considered to be 490956 kW/(m$^2$s), for those 221294 kW/(m$^2$s) are supposed to be consumed in the melting process. \ed{The total amount of heat flux is a data from the company. Moreover, we are modelling the amount of energy used for melting the material taking into account the specific heat of the sand and the amount of sand introduced in the system.}

\subsection{Numerical results and validation of the fluid flow model}

With the physical parameters defined in section 2.3 and the boundary conditions defined in section 2.4, the solution reaches the stationary state, which we define by measuring norm of the difference of temperature solution at two consecutive time steps being below $10^{-5}$. This is expected since the flow is laminar and the boundary conditions are \ed{time-independent} for times beyond $72$h, which governed the initial gradual increase of the inlet velocity. In this configuration, we reach the stationary solution at $t=188$h. \ed{The stationary state is expected to be reached since all the thermal boundary conditions are time independent, and the inlet boundary condition for velocity is constant for $t>72$ hours. Since the flux is considered to be laminar, the stationary state is reached.} In the following\ed{,} we present some results for the simulation. 

\pgfdeclareimage[width=0.75\linewidth]{TempDomain}{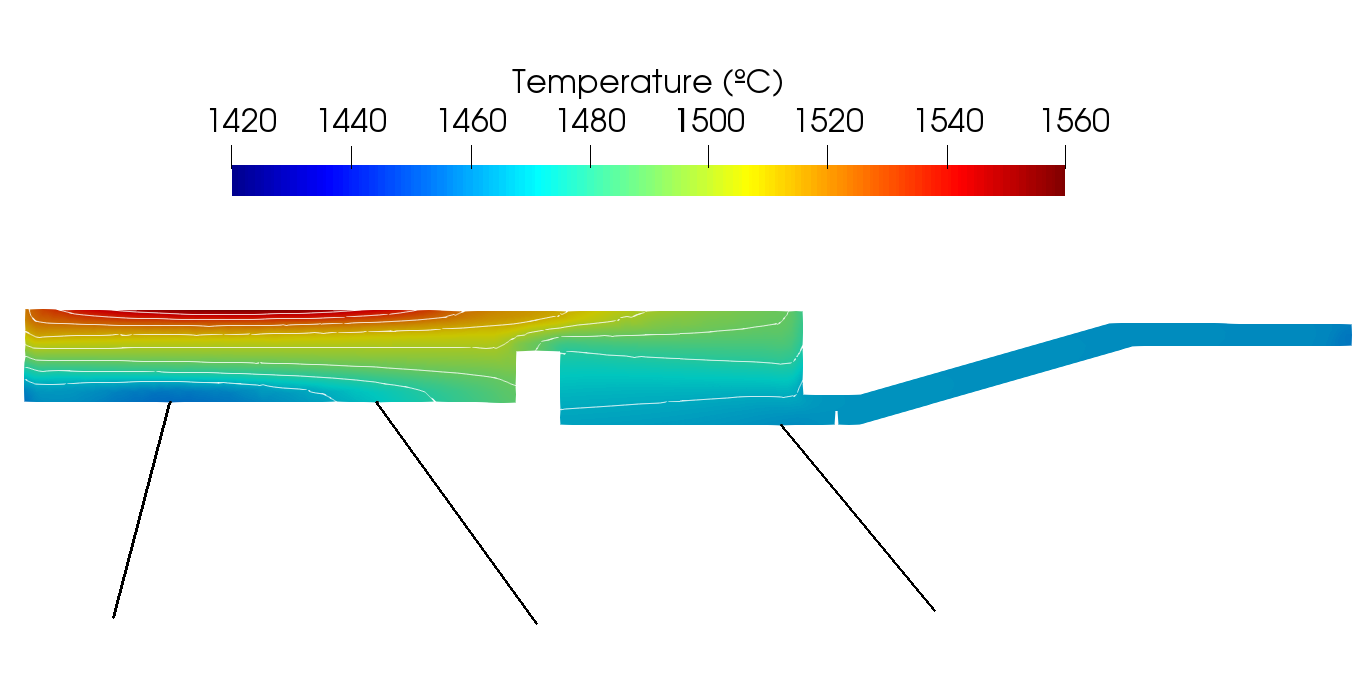}
\begin{figure}[ht]
\centering
  \begin{tikzpicture}[font=\footnotesize]
    \pgftext[at=\pgfpoint{0}{0},left,base]{\pgfuseimage{TempDomain}}
    \draw (0, 0.2) node {$T_{sim}=1454.56 \;(^{\circ}\text{C})$};
    \draw (0, -0.3) node {$T_{exp}=1454.5 \;(^{\circ}\text{C})$};
    \draw (4, 0.2) node {$T_{sim}=1470.16 \;(^{\circ}\text{C})$};
    \draw (4, -0.3) node {$T_{exp}=1499.3 \;(^{\circ}\text{C})$};
    \draw (7.5, 0.2) node {$T_{sim}=1460.7 \;(^{\circ}\text{C})$};
    \draw (7.5, -0.3) node {$T_{exp}=1399 \;(^{\circ}\text{C})$};
  \end{tikzpicture}
\caption{Comparison of simulated and measured temperatures \reviewerFour{ on a vertical plane located midway between the pairs of boosters, and aligned with the middle of the throat}.}
\label{fig:solution_points}
\end{figure}

\begin{figure}[ht]
\centering
\includegraphics[width=0.49\linewidth]{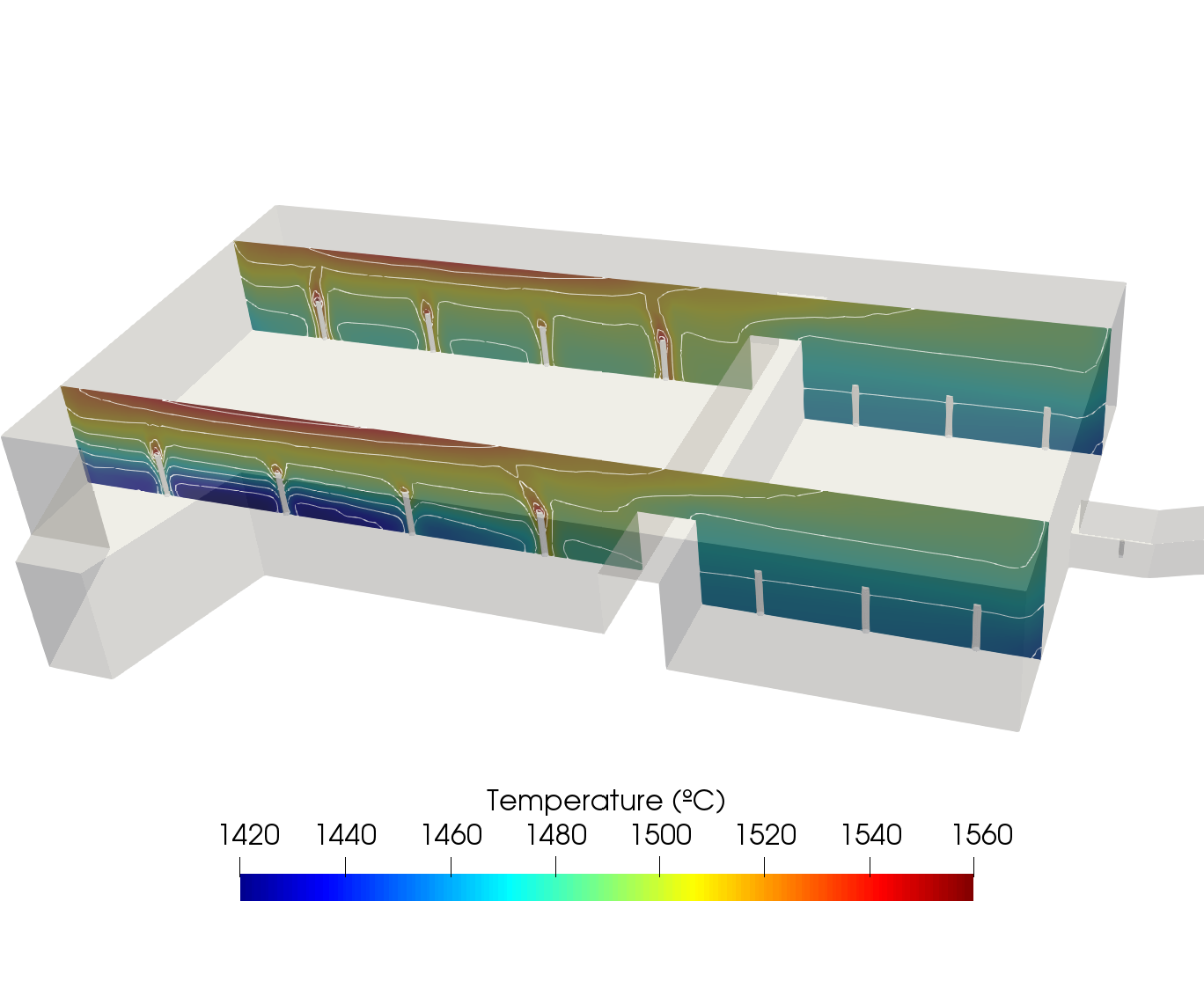}
\includegraphics[width=0.49\linewidth]{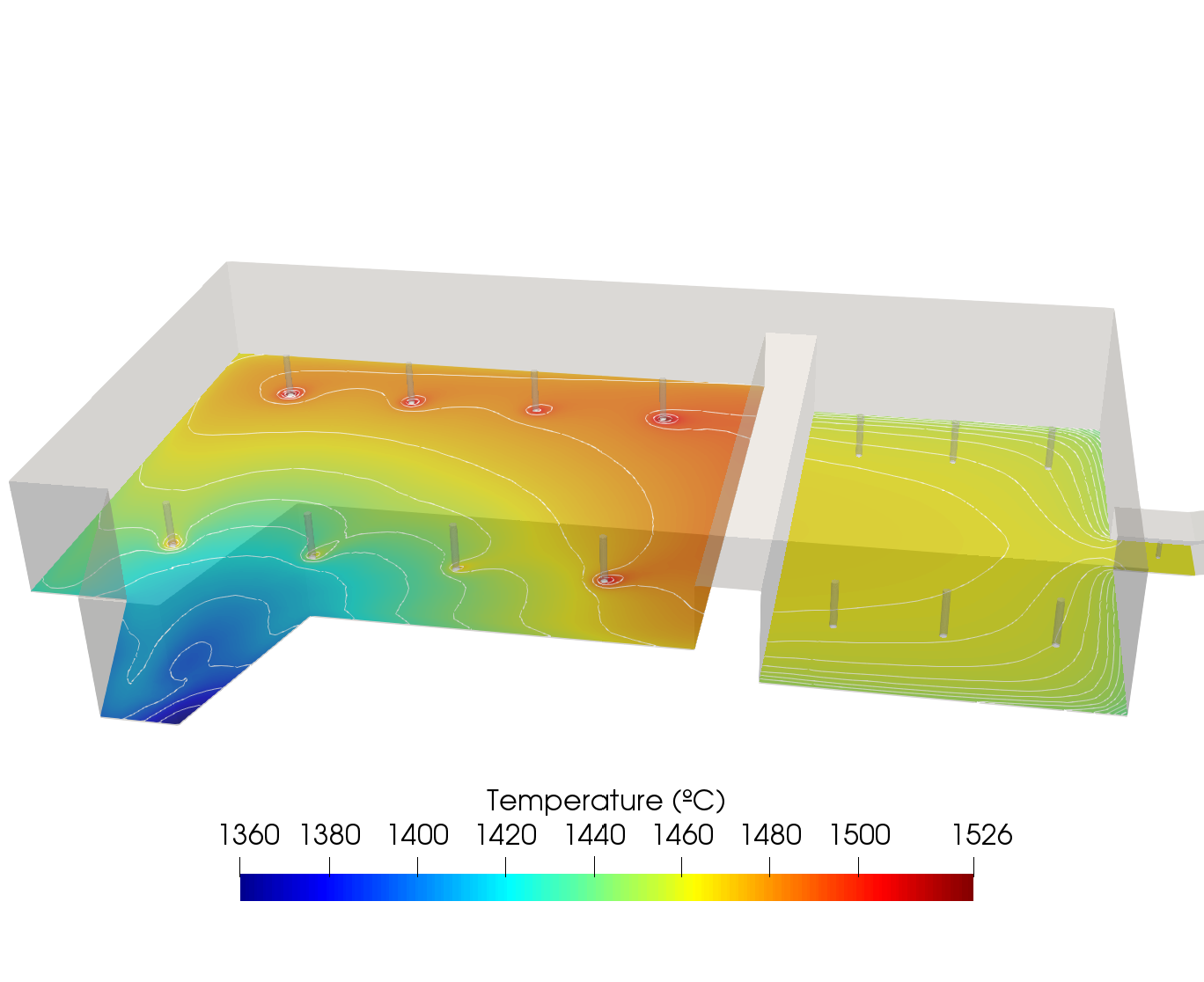}
\caption{Slides for temperature for the stationary solution.}\label{fig:Temperautres}
\end{figure}

Figure \ref{fig:solution_points} presents a comparison between temperatures obtained in the simulation $T_{sim}$, corresponding with the stationary solution, and the temperatures measured experimentally $T_{exp}$ on the vertical section of the basin which is located halfway through each pair of \ed{boosters}. \reviewerFour{Note that such a section is also the middle section of the throat --- the duct used to extract melted glass from the furnace}. We can observe how the temperatures simulated in the first part of the basin are quite similar \ed{to} the one measured experimentally, while the temperature simulated \ed{at} the end of the basin is higher than the one measured experimentally.
We can observe that there are relative errors of 0.004\%, 1.94\%\ed{,} and 4.41\%, in each point respectively.
Even the highest error, associated \ed{with} the point towards the end of the furnace, is still admissible for industrial objectives, especially considering that the physical coefficients adopted in Table~\ref{Tab:Forno2} are derived from literature, and are not fitted to the specific raw material employed in the actual experimental setting. Further slices for the temperature solution are shown in Figure \ref{fig:Temperautres}: the availability of the computational model can thus offer additional insights on the temperature distribution throughout the basin, since experimental measurements are only available on the slice reported in Figure \ref{fig:solution_points}.

\begin{figure}[ht]
\includegraphics[width=0.49\linewidth]{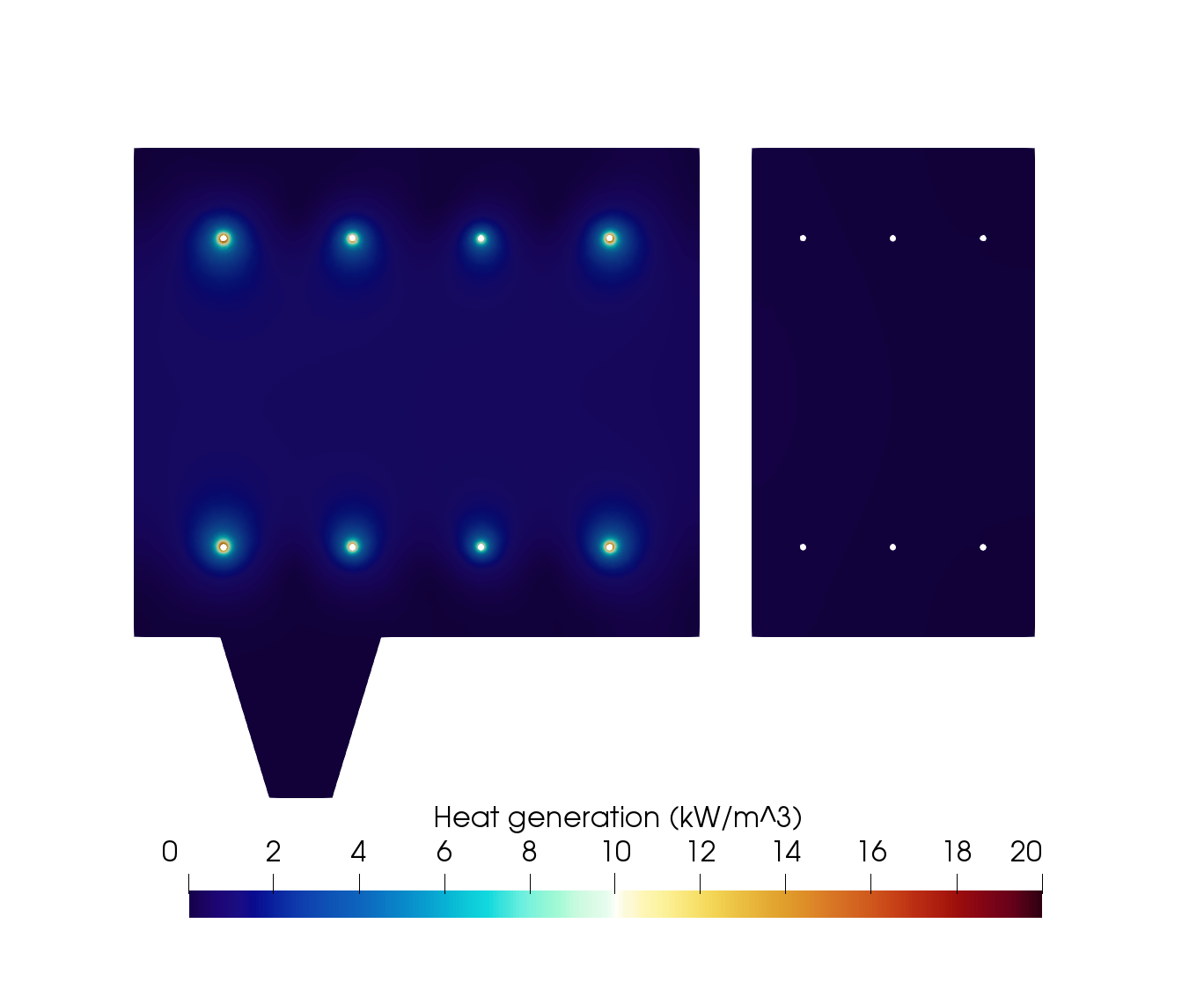}
\includegraphics[width=0.49\linewidth]{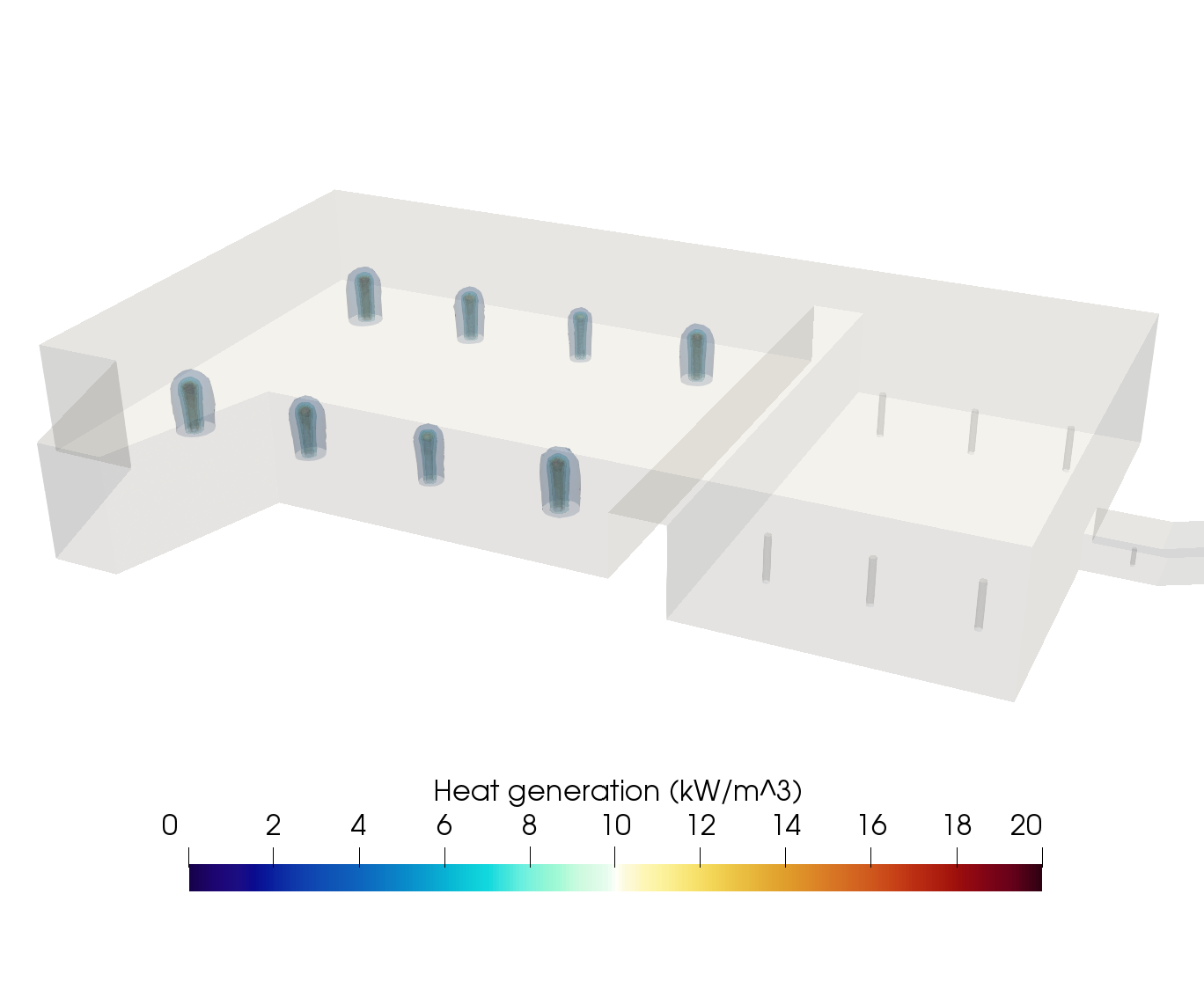}
\caption{Joule heat source for the stationary solution.}\label{fig:Slides_Joule_C1}
\end{figure}

We further show in Figure \ref{fig:Slides_Joule_C1} the values \ed{of} the Joule heat source generated by the boosters, and defined as $\sigma(\theta)|\phi|^2$. We can observe how the heat source is located basically near the \ed{boosters} that are switched on, as expected. 
 
\begin{figure}[ht]
\centering
\includegraphics[width=0.49\linewidth]{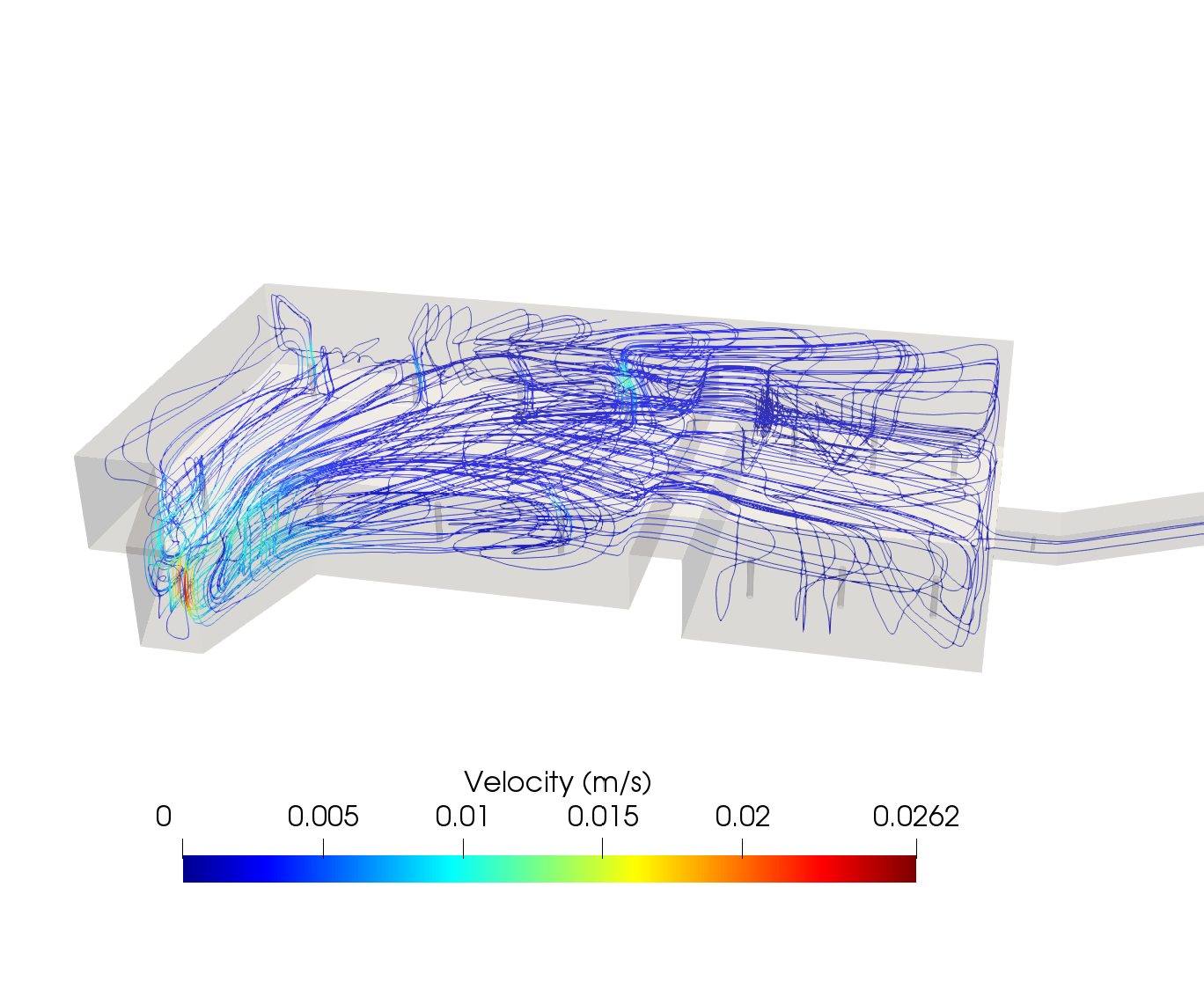}
\includegraphics[width=0.49\linewidth]{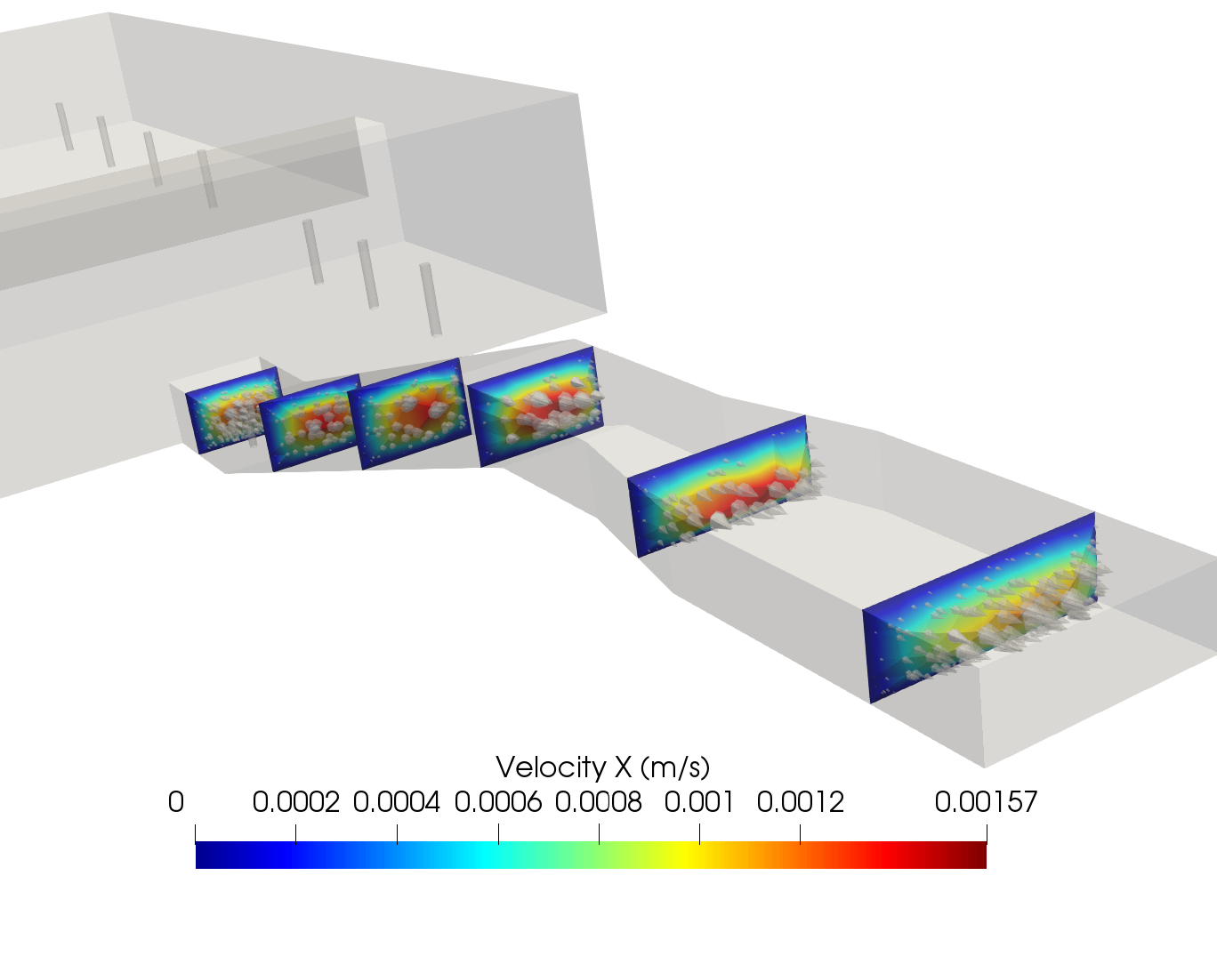}
\caption{Velocity streamlines (left) and velocity at the throat (right), for the stationary solution.}\label{fig:Streamlines_C1}
\end{figure}

For the velocity, we show some streamlines in Figure \ref{fig:Streamlines_C1} (left). As expected, we observe how the vertical velocity increases near the boosters due to the buoyancy forces. This increase of the vertical velocity is higher near the boosters that are further from the inlet. This might be explained by observing that the Joule effect heat source is higher due to the fact the temperature is higher in that zone. We also can observe a recirculation near the wall inside the bulk, that certifies the behavior expected. 

We want also to highlight the velocity profiles in the throat. In Figure \ref{fig:Streamlines_C1} (right) we show some velocity profiles in the throat, where we observe almost a parabolic profile, as expected.

\subsection{Numerical results for the air bubbles \ed{post-processing}}
We next present the numerical results of the \ed{post-processing} introduced in section 2.5 to obtain the motion of air bubbles.
In the present work, we consider bubbles having different \ed{sizes}, released in the first part of the basin, where the granular prime materials are introduced in the furnace.

\begin{figure}[ht]
\centering
\includegraphics[width=0.49\linewidth]{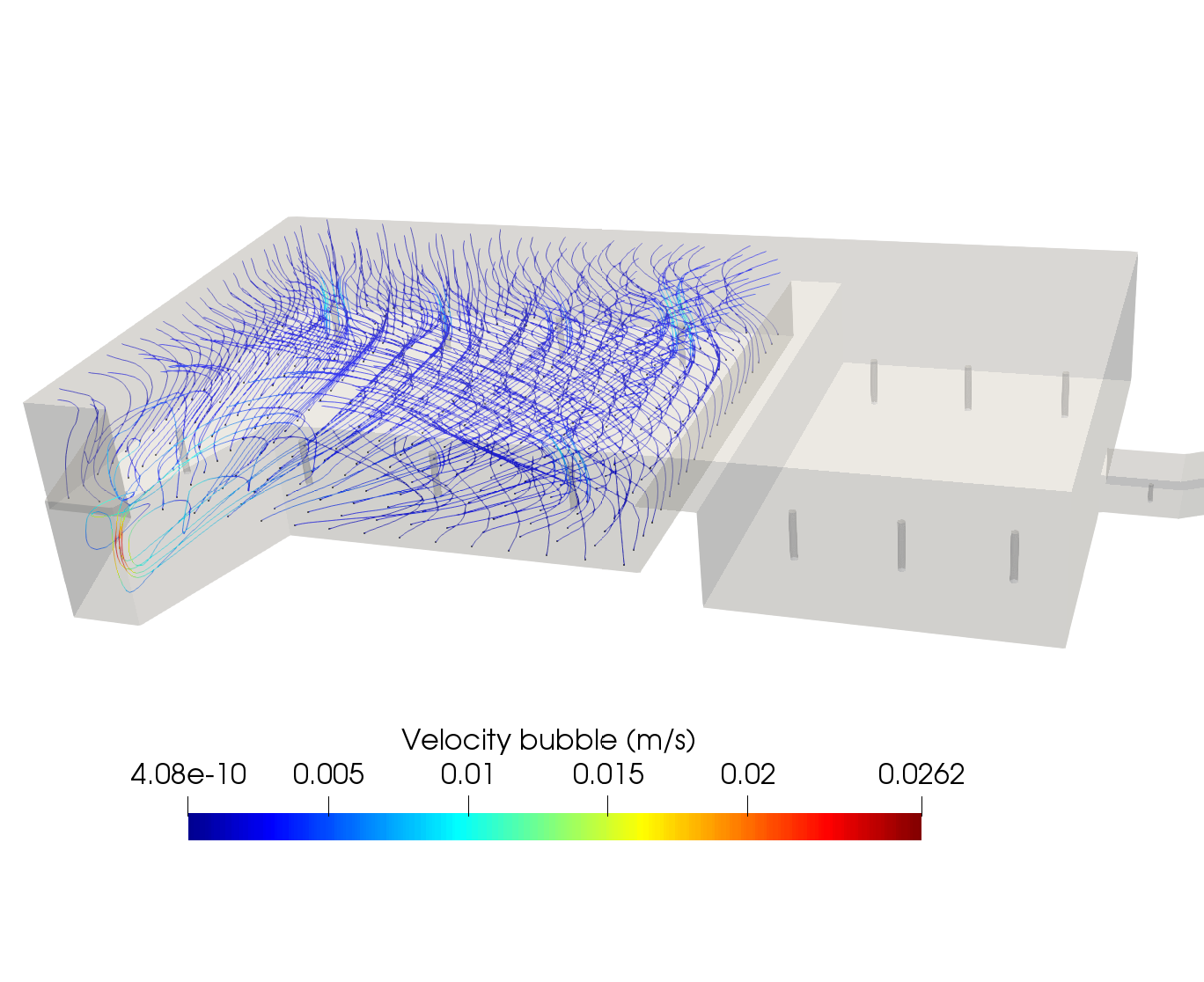}
\includegraphics[width=0.49\linewidth]{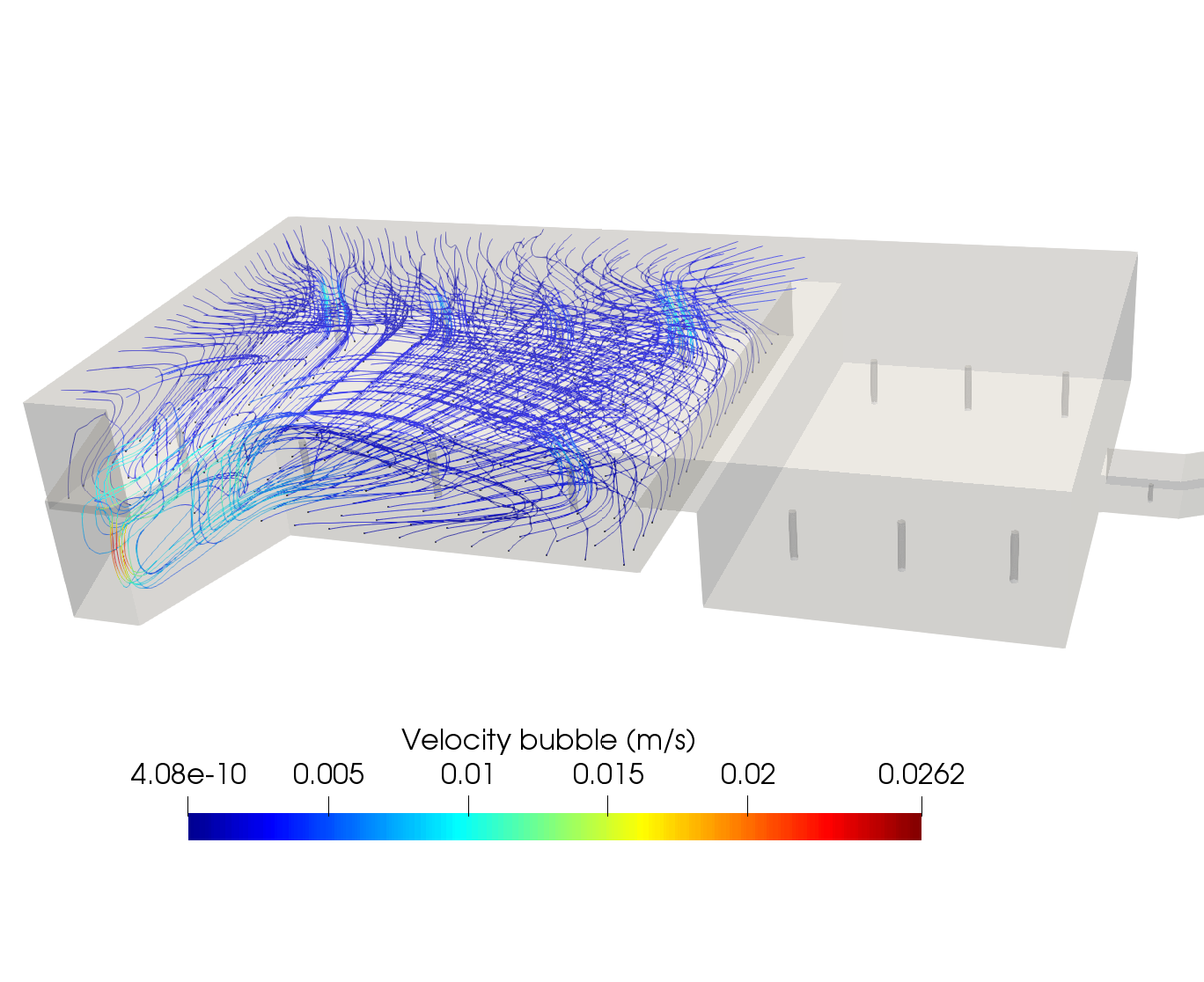}
\includegraphics[width=0.49\linewidth]{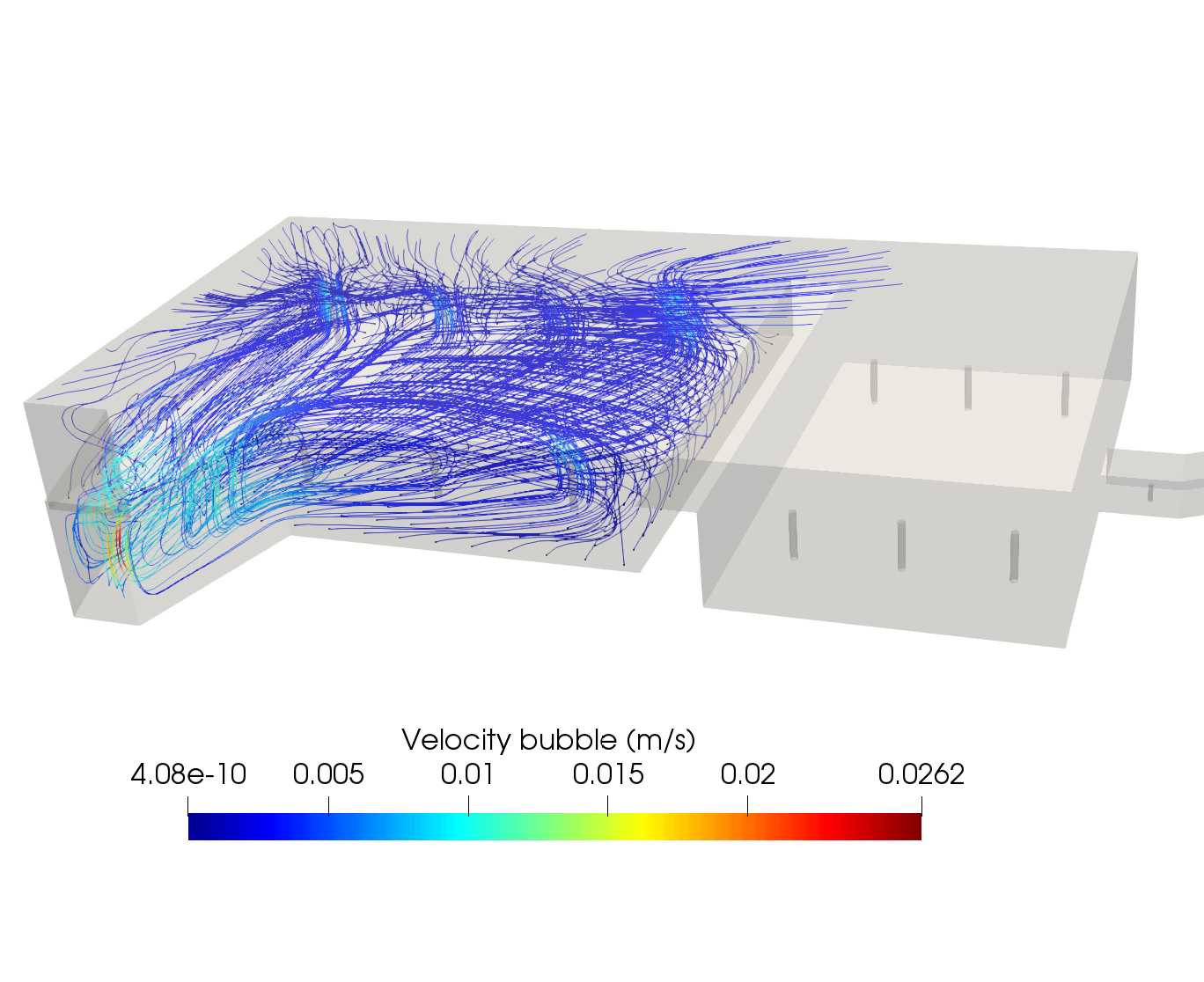}
\includegraphics[width=0.49\linewidth]{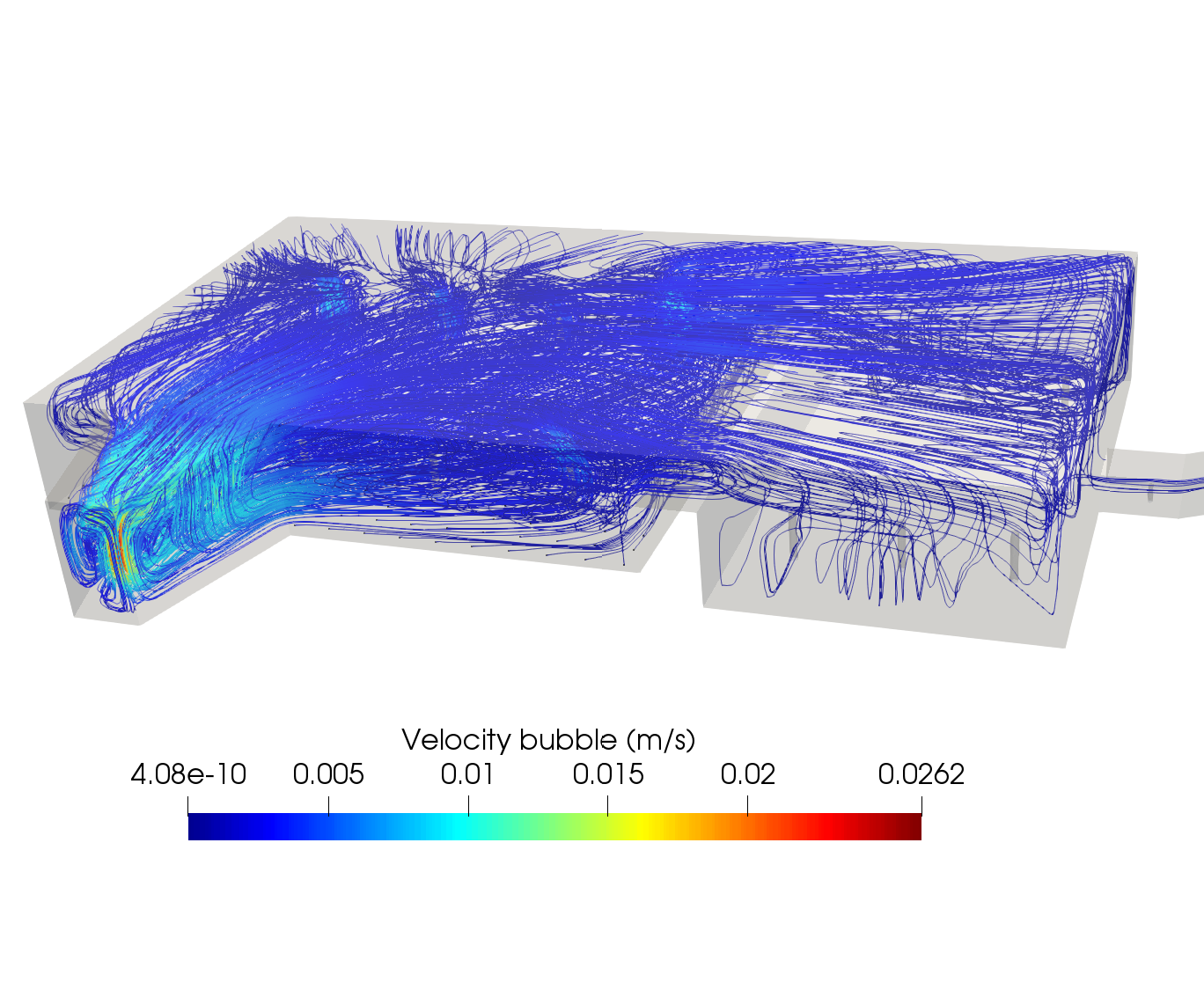}
\caption{Bubbles streamlines for diameters of 1 mm (top-left), 0.75 mm (top-right), 0.5 mm (bottom-left) and 0.25 mm (bottom-right), for the stationary solution.}
\label{fig:Bubbles}
\end{figure}

Figure \ref{fig:Bubbles} depicts pathlines of the air \ed{bubble} flow fields obtained considering different diameters. From left to right, top to bottom we show the streamlines corresponding with diameters of 1 mm, 0.75 mm, 0.5 mm and 0.25 mm, respectively. \reviewerFour{The range $[0.25\ \text{mm},1\ \text{mm}]$ has been chosen as its lower bound represents a bubble size that is --- depending on the producers --- somewhat acceptable in the finished product, and the upper bound is a size that is released relatively fast from the melting basin due to its considerable buoyancy}. All the gas spheres are released at a height of 0.1 m with respect to the furnace base. As it can be observed, practically all bubbles of 1 mm radius escape the top of the fluid. When the diameter \ed{becomes} smaller, more proportion of bubbles remain in the furnace. We can observe how, when the diameter is 0.25 mm, due to their smaller rising speed some air spheres go through the throat, remaining in the fluid at the outlet. 

The simulations and the \ed{post-processing} computation of air \ed{bubble} velocity field \ed{suggest} that the configuration studied would benefit from longer glass latency time, which would allow for smaller air bubbles to abandon the molten material, and eventually improve the final product quality.  \ed{The} operating conditions of the furnace can be changed in order to increase the glass latency time, this motivates us \ed{to introduce} parameters affecting operating conditions, and seek computational reduction as discussed in the next section.

\reviewerFour{
\subsection{POD-NN reduced order model results}
}\label{sec:ROMRes}

The reduced order model based on non-intrusive POD-ANN has been implemented in a software tool able to provide \ed{real-time} solution \ed{to} the fluid dynamic problem for each combination of parameters considered. As mentioned in Section \ref{sec:POD-NN}, the parameters considered in the simulation campaign presented in this work are related to the total amount of energy released in the furnace. Heat is in fact released in the molten glass basin both through methane combustion in the upper chamber, and through Joule effect associated with the electric currents generated by the boosters. 

Thus, in the numerical test described in the present section, we consider that the total heat flux of combustion, $q_{top}$ (see equation \eqref{eq:BC_combustion}) ranges in the interval $[240000, 300000] kW/(m^2s)$, while the voltage (see equation \eqref{eq: BC_electricity}), of each of the four pairs of boosters ranges in the interval $[100, 160] V$. With this configuration, we are assuming that our parametrized PDE depend on five parameters, which vary in the parameter space $\mathcal{D} = [240000, 300000] \times [100, 160]^4$.
\reviewerFour{A total of 200 FOM simulations have been carried out so as to assemble and validate the POD-NN model described. The results of the first 150 snapshots were obtained with parameters in the training set, namely $\mu\in\cD_{train}$. These solutions were first used to produce the snapshot matrix from which the POD modes were extracted, and then train the feed-forward Neural Network used to compute the modal coefficients at the reduced order level. The residual 50 solutions computed, associated with the testing parameters $\mu\in\cD_{test}$  have then been used for validation purposes.}

\ed{We have computed the FOM solutions in a cluster with IBM nodes: Xeon E5-2680 v2 of 40GB of RAM. We have computed the FOM solution with 40 processors, and each FOM solution takes about 3 hours to be computed  each one. Thus, to perform the offline phase (FOM computation plus training of the Neural Network), it took about one month to be completed. Then, the online phase, as can be viewed at the website \url{https://youtu.be/yRNQaGVmXF4}, it is performed on real time.}

\begin{figure}[htbp]
    \centering
    \includegraphics[width=0.75\linewidth]{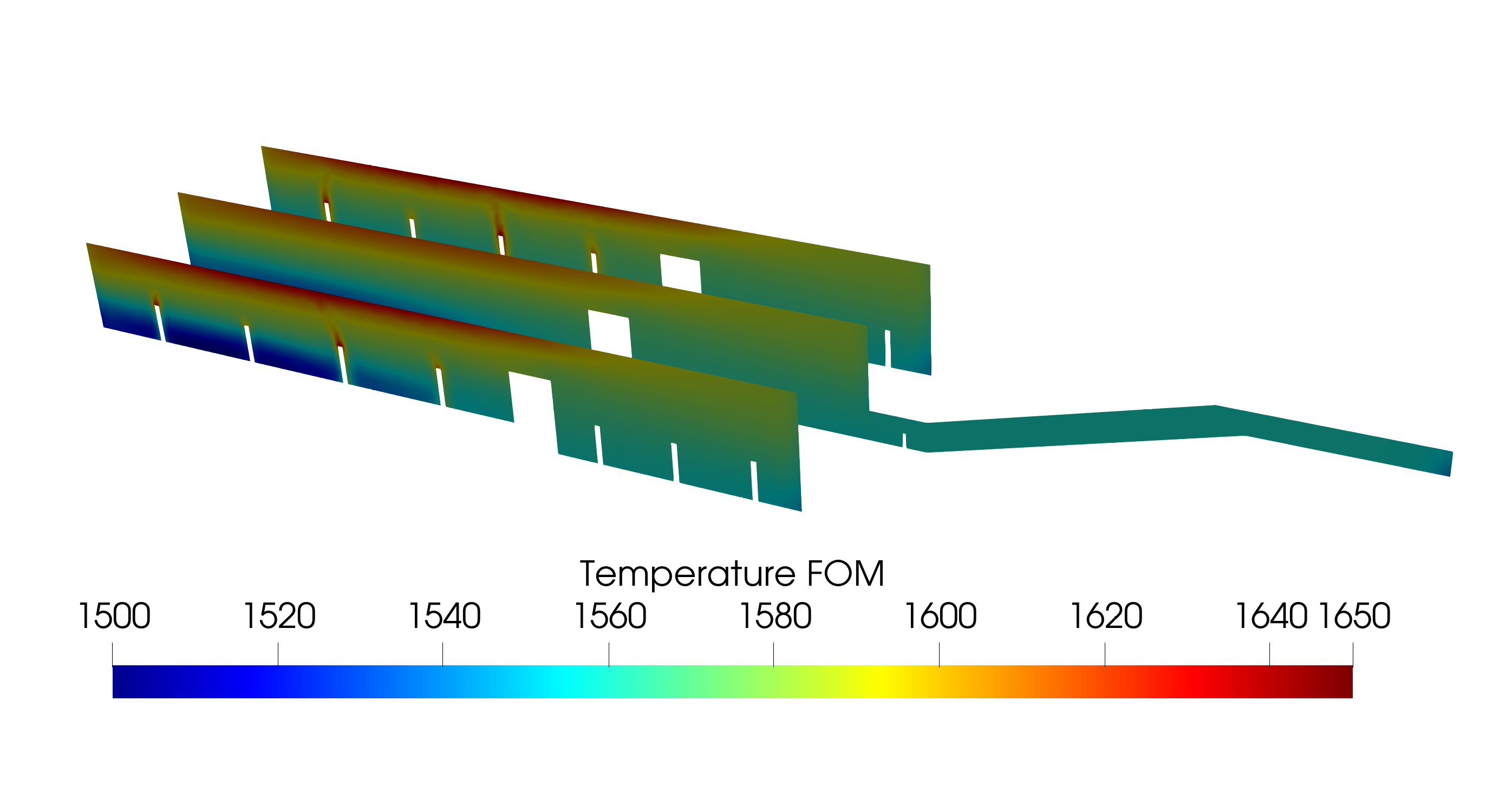}
    \includegraphics[width=0.75\linewidth]{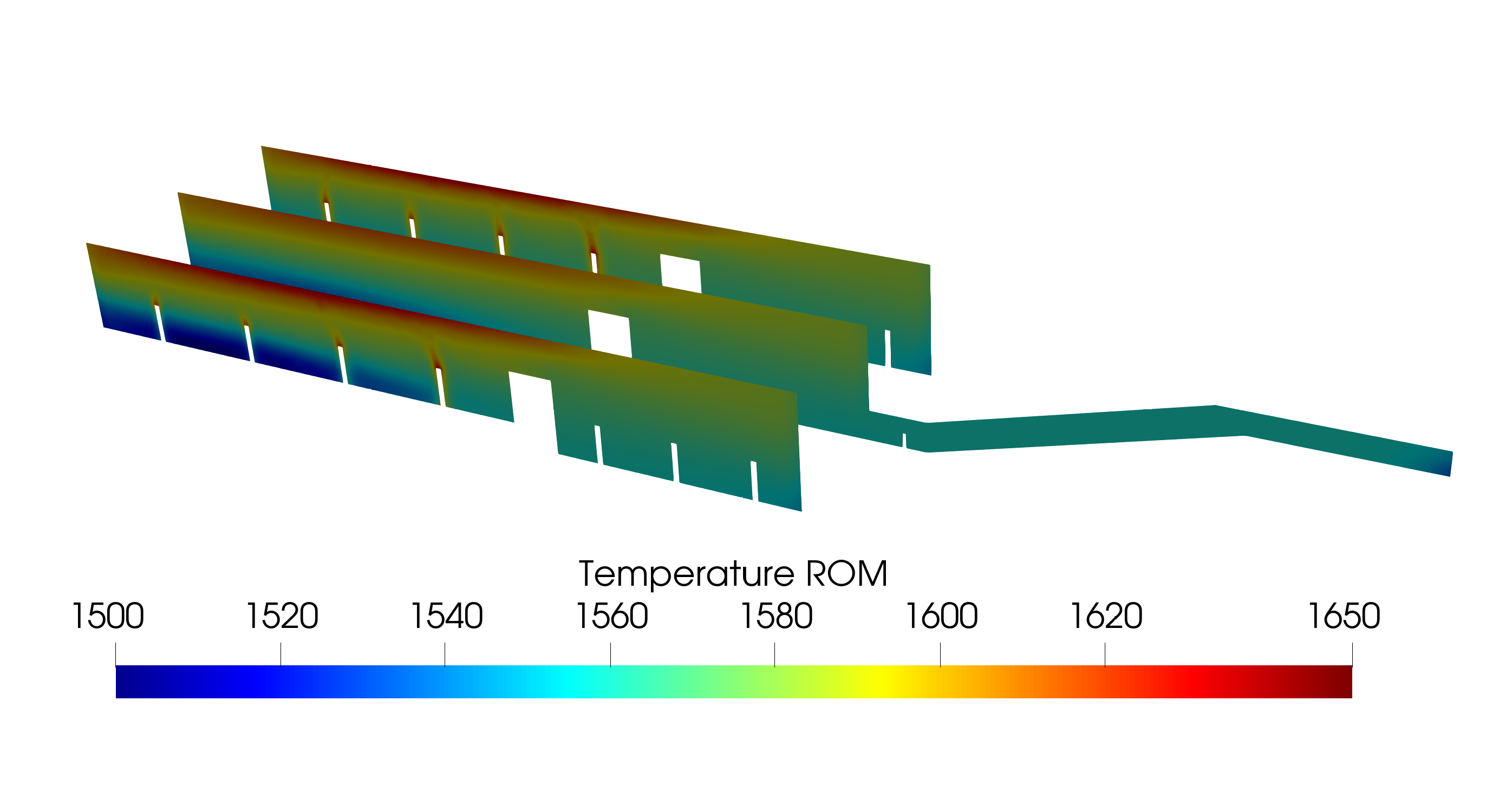}
    \includegraphics[width=0.75\linewidth]{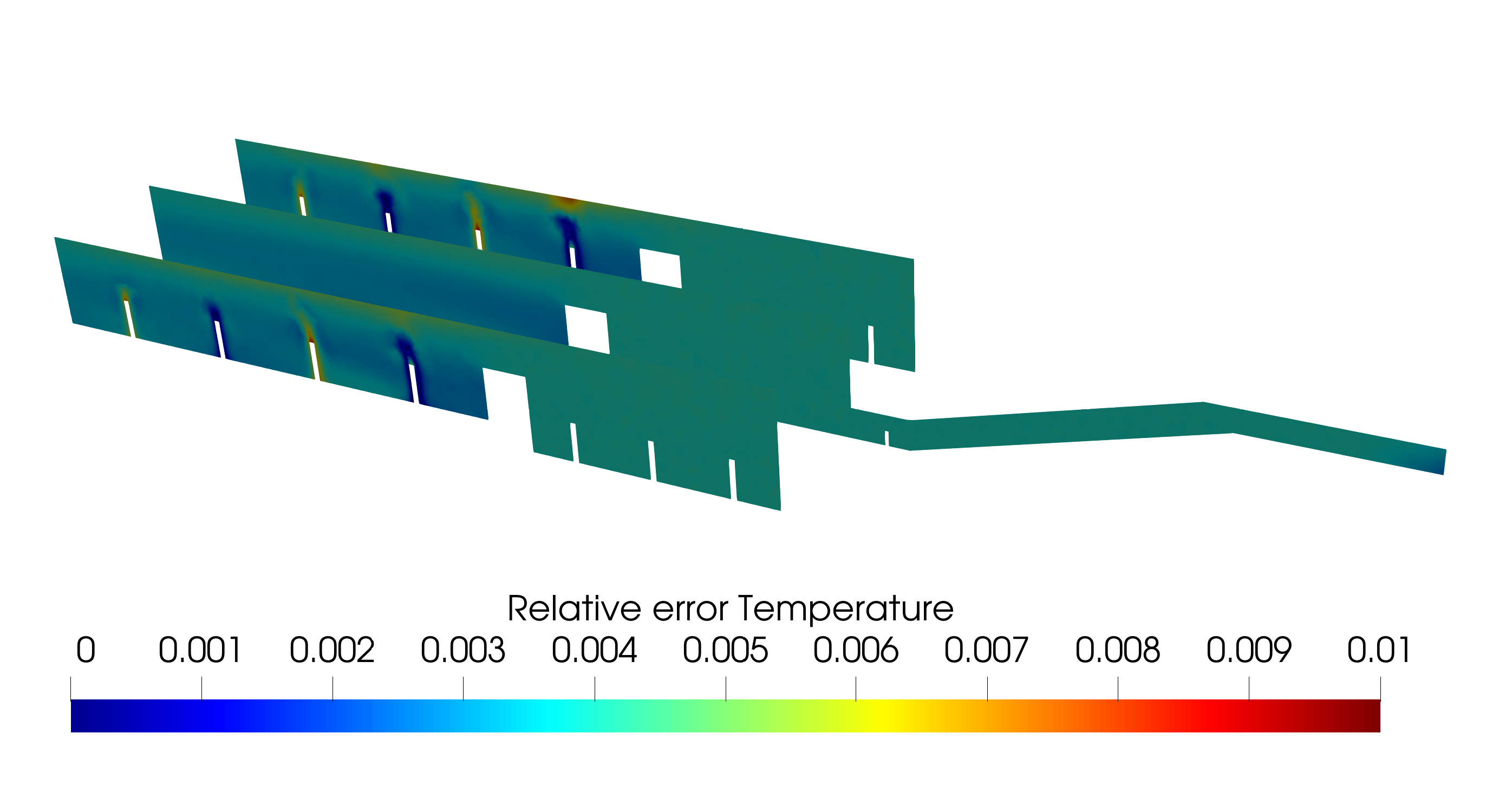}
    \caption{From top to bottom: Temperature of FOM solution, temperature of ROM solution, and the relative error between FOM and ROM of the temperature field. An animation of the solution variations resulting from modification of the value of different process parameters can be viewed at the website \url{https://youtu.be/yRNQaGVmXF4}.}
    \label{fig:FOM_ROM_Temp}
\end{figure}

Figure \ref{fig:FOM_ROM_Temp} displays a comparison between the ROM and FOM solutions obtained for a parameter setting corresponding to $283961.719kW/(m^2/s)$ of combustion heat flux, and $146.681V$, $116.542V$, $150.134V$, and $106.326V$ for the voltage of each pair of booster, respectively. The plot on the top depicts contours of temperature field predicted with the FOM solver, on three vertical planes within melting the basin. Using an identical layout, the middle plot illustrates the \ed{the contours of the temperature field} computed by means of the ROM algorithm. Finally, the bottom plot depicts \ed{the} contours of the relative error between ROM and FOM temperature on the three vertical planes used in the temperature plots.  We can observe the similarity of both FOM and ROM solution. Actually, in this case, the maximum of the local relative error of both solutions is about $1\%$. 

\begin{figure}[htbp]
    \centering
    \includegraphics[width=\linewidth]{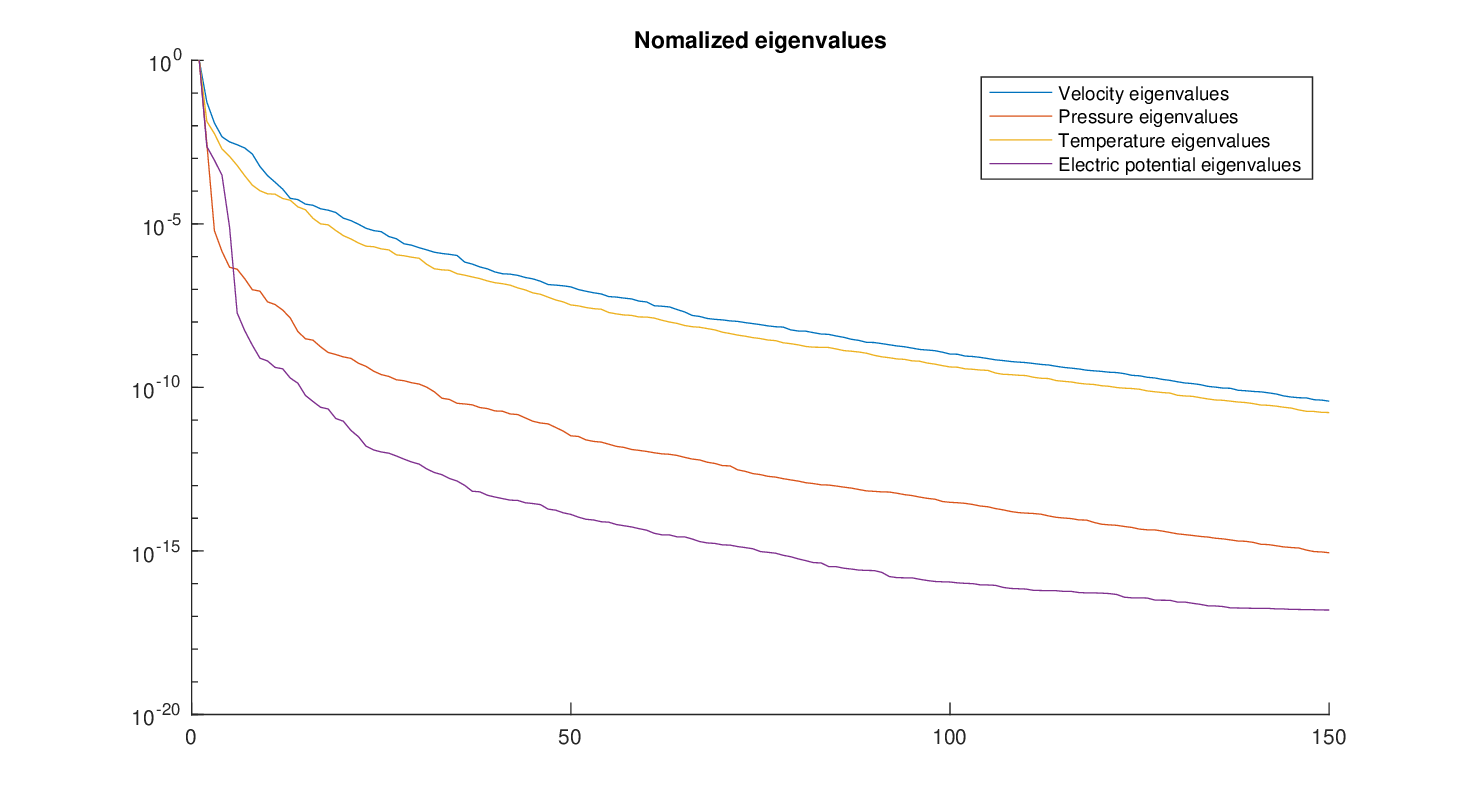}
    \caption{Normalized eigenvalues for velocity, pressure, temperature and electric potential}
    \label{fig:POD_eigen}
\end{figure}

In addition, we compute the overall POD-NN relative error, $e_\uk, e_p, e_\theta, e_\phi$, of each field for a given parameter value $\muk\in\cD$, as
\[
    e_\uk(\muk) = \dfrac{\|\uk_h(\muk) - \uk_{NN}(\muk)\|}{\|\uk_h(\muk)\|},
    \quad
    e_p(\muk) = \dfrac{\|p_h(\muk) - p_{NN}(\muk)\|}{\|p_h(\muk)\|},
\]
\[
    e_\theta(\muk) = \dfrac{\|\theta_h(\muk) - \theta_{NN}(\muk)\|}{\|\theta_h(\muk)\|},
    \quad
    e_\phi(\muk) = \dfrac{\|\phi_h(\muk) - \phi_{NN}(\muk)\|}{\|\phi_h(\muk)\|}.
\]

\begin{figure}[htbp]
    \centering
    \includegraphics[width=\linewidth]{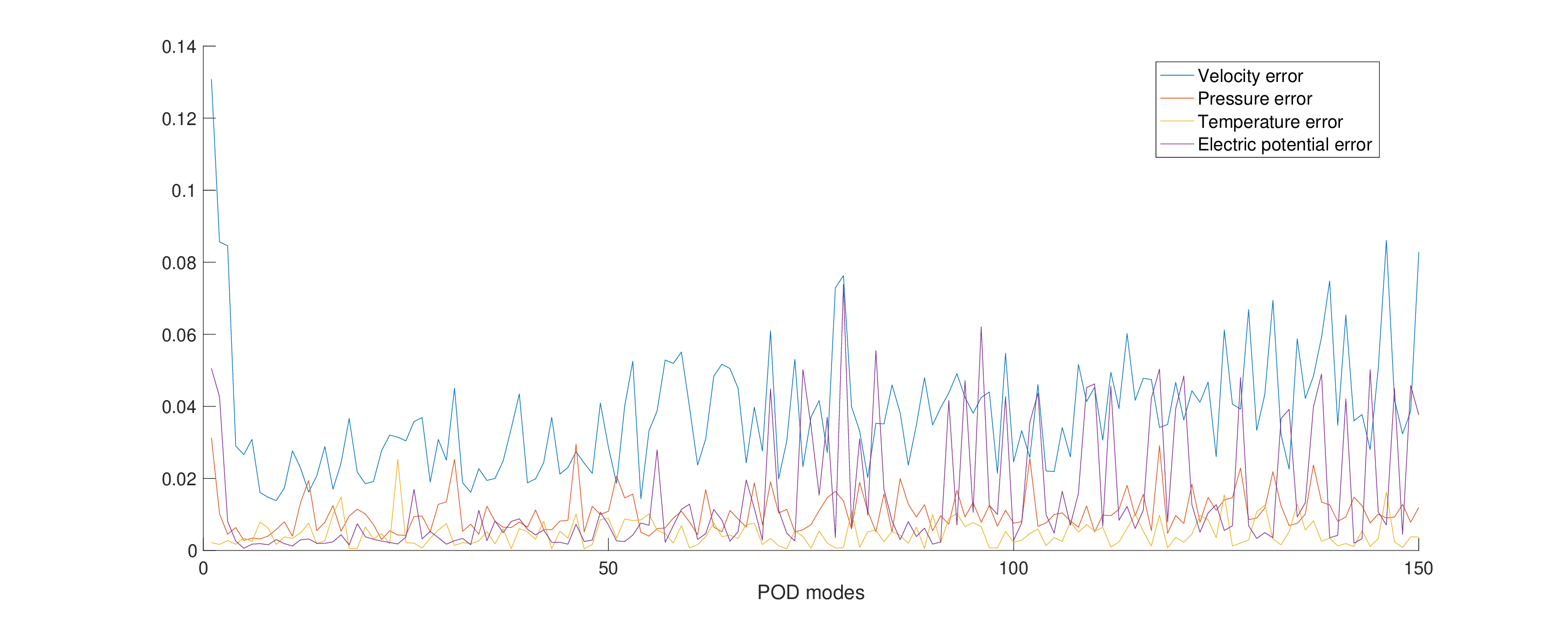}
    \caption{Comparative of relative errors for velocity, pressure, temperature and electric potential.}
    \label{fig:NN_error_together}
\end{figure}

We consider a set of parameter values, $\cD_{test}$, for which we compute the POD-NN solution relative error with respect to the full order solutions. In Figure \ref{fig:NN_error_together}, we show the mean of the POD-NN relative errors for velocity, pressure, temperature and electric potential, depending on the number of POD modes we have considered to construct the reduced spaces. 
The plot shows that, among the variables considered, velocity is the one comparing the order of magnitude of them.


We can observe that the POD-NN relative error, with respect to the high fidelity solution, ranges between $6\%$ to $1\%$, noticing that the error in temperature is the lowest one. We can observe also a high variability on the mean of the relative error depending on the number of POD modes. That means that we do not need a high number of POD modes in order to have a good approximation, with respect to the high fidelity solution, with the POD-NN approach considered. 

\reviewerFour{In addition, the plot clearly shows that the POD-NN approximation error with respect to the FOM solutions is growing as the number of modes considered in the ROM increases. Of course, this might seem a counter intuitive occurrence, as the error between ROM and FOM is ideally expected to be reduced by considering more modes in the ROM. However, we must remark that in this work we only had 200 FOM solutions available, 150 of which were used for training. Thus, as we increased the number of modes considered, and as the dimension of the NN output space was growing, we could not increase accordingly the amount of samples used for training. In light of this consideration, the observed degradation of the ROM accuracy has to be ascribed to a loss of performance of the NN training. \ed{In Figure \ref{fig:POD_eigen} we show the decay of the normalized eigenvalues for each field. We can observe how at the beginning, the decay is quite fast, while for the last values this decay get slower. Moreover, the decay rate of each field is quite similar. Since the main decay of the eigenvalues is produces for the first nodes, it could be that there is when the performance of the NN training is lost.} Clearly, we could have increased the offline computational effort and provide the Neural Network with additional training samples and snapshots to produce decreasing error curves, as the ones observed for the first 10 modes. 

However, we feel that having a cap on the number of full order simulations is a more realistic scenario in real world applications. In fact, in most situations the additional computational time --- and money --- required for each FOM simulation, and even the amount of storage required to save snapshots for each field considered in the PDE solution, put a hard limit to the amount of FOM simulations that can be used for training. A plot like that presented in Figure \ref{fig:NN_error_together} is then representing a relatively common situation in the realm of industrial applications.} \\

\section{Conclusions}

In this work we have presented the modelling of a 3D flow of a molten glass fluid inside a furnace. We have presented the numerical results of a FEM simulation and compared them with experimental data. We have also modelled the behaviour of air bubbles inside the fluid depending on their size. \ed{Moreover, we have presented a POD-NN model in order to reduce the computational time of the numerical solution. }
We have performed this POD-NN model with different number of POD modes and we have compared the solution obtained from the non-intrusive POD-NN model with respect to the FE solution. For all the fields, we have obtained an error ranging between $6\%$ and $1\%$, the error in temperature being the lowest one. 
\ed{Also, we have seen that that the consideration of non-intrusive ROM for industrial problems can provide acurrate approximations of the solution for a parametrized setting. As future work, we intend to consider alternative non-intrusive methods, such as a POD-RBF, in order to recover the POD coefficients. In order to improve the Neural Network training, we should need to increase the offline computational effort and provide the Neural Network with additional training samples and snapshots to produce decreasing error curves in the POD-NN model.}

\section*{Acknowledgments}
We acknowledge the European Union's Horizon 2020 research and innovation program under the Marie Skłodowska-Curie Actions, grant agreement 872442 (ARIA).
Francesco Ballarin acknowledges the INdAM-GNCS project ``Metodi numerici per lo studio di strutture geometriche parametriche complesse'' (CUP E53C22001930001), the project ``Reduced order modelling for numerical simulation of partial differential equations'' funded by Università Cattolica del Sacro Cuore and the PRIN 2022 PNRR project ``ROMEU: Reduced Order Models for Environmental and Urban flow'' (CUP J53D23015960001) funded by MUR. Andrea Mola also acknowledges the funding by project ROMEU (CUP D53D2301888001).  
\section*{Conflict of interest}
Not Applicable

\bibliography{Glass_references.bib}

\end{document}

%% file: Figure/boundaries.pdftex_t
\begin{picture}(0,0)%
\includegraphics{Figure/boundaries.pdf}%
\end{picture}%
\setlength{\unitlength}{4144sp}%
\begingroup\makeatletter\ifx\SetFigFont\undefined%
\gdef\SetFigFont#1#2#3#4#5{%
  \reset@font\fontsize{#1}{#2pt}%
  \fontfamily{#3}\fontseries{#4}\fontshape{#5}%
  \selectfont}%
\fi\endgroup%
\begin{picture}(8655,5985)(2101,-7711)
\put(2791,-5821){\makebox(0,0)[lb]{\smash{{\SetFigFont{12}{14.4}{\familydefault}{\mddefault}{\updefault}{\color[rgb]{0,0,0}$\Gamma_{in}$}%
}}}}
\put(4231,-6766){\makebox(0,0)[lb]{\smash{{\SetFigFont{12}{14.4}{\familydefault}{\mddefault}{\updefault}{\color[rgb]{0,0,0}$\Gamma_{w}$}%
}}}}
\put(6121,-6496){\makebox(0,0)[lb]{\smash{{\SetFigFont{12}{14.4}{\familydefault}{\mddefault}{\updefault}{\color[rgb]{0,0,0}$\Gamma_{wr}$}%
}}}}
\put(6616,-7306){\makebox(0,0)[lb]{\smash{{\SetFigFont{12}{14.4}{\familydefault}{\mddefault}{\updefault}{\color[rgb]{0,0,0}$\Gamma_{in}$}%
}}}}
\put(9046,-4831){\makebox(0,0)[lb]{\smash{{\SetFigFont{12}{14.4}{\familydefault}{\mddefault}{\updefault}{\color[rgb]{0,0,0}$\Gamma_{duct}$}%
}}}}
\put(9856,-4246){\makebox(0,0)[lb]{\smash{{\SetFigFont{12}{14.4}{\familydefault}{\mddefault}{\updefault}{\color[rgb]{0,0,0}$\Gamma_{b_i}$}%
}}}}
\put(9136,-2536){\makebox(0,0)[lb]{\smash{{\SetFigFont{12}{14.4}{\familydefault}{\mddefault}{\updefault}{\color[rgb]{0,0,0}$\Gamma_{top}$}%
}}}}
\put(7696,-2086){\makebox(0,0)[lb]{\smash{{\SetFigFont{12}{14.4}{\familydefault}{\mddefault}{\updefault}{\color[rgb]{0,0,0}$\Gamma_{bott}$}%
}}}}
\put(2937,-7638){\makebox(0,0)[lb]{\smash{{\SetFigFont{12}{14.4}{\familydefault}{\mddefault}{\updefault}{\color[rgb]{0,0,0}$x$}%
}}}}
\put(2605,-6380){\makebox(0,0)[lb]{\smash{{\SetFigFont{12}{14.4}{\familydefault}{\mddefault}{\updefault}{\color[rgb]{0,0,0}$y$}%
}}}}
\put(2116,-7261){\makebox(0,0)[lb]{\smash{{\SetFigFont{12}{14.4}{\familydefault}{\mddefault}{\updefault}{\color[rgb]{0,0,0}$z$}%
}}}}
\end{picture}%